\documentclass{pinchcr}
\usepackage{graphicx,amsmath}
\usepackage{epstopdf, epsfig}
\newcommand{\cal}{\mathcal}

\newcommand{\mathbb}{\tilde}
\renewcommand{\v}{{\boldsymbol{v}}}

\newcommand{\f}{{\boldsymbol{f}}}
\newcommand{\0}{{\boldsymbol{0}}}
\newcommand{\h}{{\boldsymbol{h}}}
\newcommand{\D}{{\boldsymbol{D}}}
\newcommand{\nuhat}{{\hat{\boldsymbol{\nu}}}}

\renewcommand{\u}{{\boldsymbol{u}}}

\newcommand\bdot{{\boldsymbol{\cdot}}}
\newcommand\del{{\boldsymbol{\nabla}}}
\newcommand\delsq {\nabla^2}
\newcommand\bcol{{\boldsymbol{:}}}
\renewcommand{\r}{{\boldsymbol{r}}}

\newcommand{\J}{{\boldsymbol{J}}}
\renewcommand{\d}{{\rm d}}
\newcommand{\p}{{\boldsymbol{p}}}
\newcommand{\bsigma}{{\boldsymbol{\sigma}}}
\newcommand{\bSigma}{{\boldsymbol{\Sigma}}}
\newcommand{\bepsilon}{{\boldsymbol{\epsilon}}}
\newcommand{\bLambda}{{\mbox{\boldmath{$\Lambda$}}}}

\newcommand{\Q}{{\boldsymbol{Q}}}
\newcommand{\A}{{\boldsymbol{A}}}
\renewcommand{\H}{{\boldsymbol{H}}}
\newcommand{\V}{{\boldsymbol{V}}}
\newcommand{\bOmega}{{\boldsymbol{\Omega}}}
\newcommand{\I}{{\boldsymbol{I}}}
\renewcommand{\del}{\nabla}
\renewcommand{\bdot}{\cdot}
\DeclareMathOperator{\Tr}{Tr}

\title{Active Field Theories}
\subtitle{Lecture Notes for les Houches 2018 Summer School on Active Matter and Nonequilibrium Statistical Physics}
\author{Michael E. Cates}
\affiliation{DAMTP, University of Cambridge, Centre for Mathematical Sciences, Wilberforce Road, Cambridge CB3 0WA, United Kingdom}
\begin{document}
\maketitle
\tableofcontents
\maintext
%
\chapter{Active Field Theories}\label{ch:one}
\section{Field theories in soft matter}
Within the realm of soft condensed matter, dynamical field theories play a pivotal role. These continuum theories are developed by constructing a coarse grained description involving a set of order parameters $\Psi(\r, t)$, whose evolution equations can be found either by explicit coarse-graining of microscopic models (the bottom-up approach), or by top-down considerations. In the latter,  the leading-order terms  --- in an expansion in powers of $\Psi$ and its spatial and temporal derivatives --- are presumed to all enter with unknown coefficients, except for those terms disallowed by symmetries and conservation laws for the system in question. For overdamped systems of the type considered in these notes, one has typically a (set of) first order PDE(s) as the equation(s) of motion ($\dot\Psi(\r,t) = ...$) whose deterministic version is often referred to as a `hydrodynamic' description. This controls the mean behaviour of the system, but we are often also interested in its fluctuation behaviour, in which case noise must be added to create a model at stochastic PDE level. Such PDEs governs the `path probabilities' ${\mathbb P}[\Psi(\r,t)]$ for different spatiotemporal evolutions of the system. (We use this notation to distinguish a path probability from a configurational probability $P[\Psi(\r)]$.)

For passive materials, in which there is no internal driving of the particles (requiring a continuous feed of energy), nor global driving such as a shear flow, the behaviour of $\Psi(\r,t)$ is strongly constrained by the requirement that the unique steady state probability distribution $P[\Psi(\r)]$, for configurations $\Psi$, is the Boltzmann distribution, $P_B \propto \exp[-\beta F[\Psi]]$. Recall that this is controlled by the Helmholtz free energy $F$ at coarse-grained level; $F$ becomes the Hamiltonian $H$ if all microscopic details are retained~\cite{Chaikin95}. In general, this requirement is enough to fix the noise terms in the equations of motion for $\Psi$ via the fluctuation dissipation theorem (an explicit example is shown later). For active matter, there can be no similar appeal to the Boltzmann distribution and noise terms have to be either found by explicit coarse-graining, or simply guessed. 

The rest of this introductory Section will survey classical field theories for passive soft materials such as interacting Brownian particles, binary fluid mixtures, and liquid crystals. In Section~\ref{ch:two} I will give a similar overview of some field theories used to describe active materials. The remaining sections specialize to those active systems whose large-scale behaviour can be adequately described using a single scalar order parameter governing, for instance, the density of active particles (possibly coupled to a fluid velocity). Section~\ref{ch:three} constructs a field theory for a specific model in that class, by explicit coarse-graining of its microscopic dynamics, and explores the resulting phase behaviour at mean-field level. This phase behaviour captures the phenomenology of `MIPS', or motility-induced phase separation. Section~\ref{ch:four} presents a simplified `canonical' version of the same model (called Active Model B) which directly connects with top-down approaches. Section~\ref{ch:five} briefly describes the extension of this model to the case where our active scalar is coupled to a momentum conserving solvent (Active Model H), necessitating the consideration of a fluid velocity field alongside the scalar density. Section~\ref{ch:six} addresses the absence within Active Model B of certain terms that, even if less easily motivated microscopically, should really be present in a top-down field theory for MIPS. These terms, combined with the presence of noise, radically alter the phase behaviour, allowing the appearance of microphase separated states in which bulk phase separation is curtailed at a certain characteristic length scale. This offers a general explanation for states involving a set of finite-sized vapour droplets surrounded by liquid, or finite-sized dense clusters surrounded by vapour, that have been reported in some active systems both experimentally and numerically. Section~\ref{ch:seven} contains a brief conclusion and outlook.

\subsection{Order parameters for soft matter}\label{OPs}
In soft matter, a few of the order parameters $\Psi(\r,t)$ that may be needed --- depending on the system ---  are as follows. (i) A scalar density $\rho$ in a single-component (compressible fluid). (ii) A fluid velocity $\v$. (iii) A scalar composition variable $\phi$ in a two-component mixture. (iv) A polarization vector $\p$ describing the orientation of polar particles (possibly oriented swimmers). (v) A second-rank symmetric traceless tensor $\Q$ describing the nematic order of rodlike particles that orient along an axis but with no preferred sense of direction along that axis. All will feature in later examples. 

Consider first the case of a single-component, incompressible fluid. Here $\rho$ is not an order parameter (it is constant in space and time) but just a parameter of the model. The hydrodynamic equation for such a fluid, which we assume to be isothermal, is the familiar Navier Stokes equation (NSE),
\begin{equation}
\rho(\dot \v + \v\bdot\del\v) = \eta\delsq \v - \del P , \label{NSE}
\end{equation}
where $\eta$ is the visosity and the pressure field $P$ must be chosen to enforce the incompressibility condition 
\begin{equation}
\del\bdot\v = 0 .\label{continuity}
\end{equation}
Note that in the absence of forcing (by body forces or at boundaries) the only steady-state solution of the NSE is $\v = \0$. 
This is not the Boltzmann distribution for a fluctuating thermal velocity field ($P[\v] \propto \exp[-\beta \rho \int|\v|^2d\r/2]$) and leads to a description where, for instance, Brownian motion of a suspended colloidal particle is absent.

To correct this we add a noise term which can be determined using the fluctuation dissipation theorem (FDT). This takes the form $\del\bdot\bSigma^N$ where the random stress tensor is a zero-mean Gaussian quantity whose statistics obey~\cite{Landau59}
\begin{equation}
\langle \Sigma^N_{ij}(\r,t)\Sigma^N_{kl}(\r',t')\rangle = 2k_BT\eta[\delta_{ik}\delta_{jl}+\delta_{il}\delta_{jk}]\delta(\r-\r')\delta(t-t').\label{Hvnoise}
\end{equation}
Note that due to incompressibility the isotropic (pressure-like) part of this noise stress is optional, and often explicitly removed in the literature. The reason that the noise takes the form of a stress is that the NSE expresses conservation of momentum; the whole of the right hand side can therefore be written as the divergence of a momentum flux (or stress tensor) comprising viscous, isotropic and now noisy contributions.

While it is not commonplace to refer to the deterministic NSE as a field theory, the stochastic version is a good example of one. 
A second example comprises interacting Brownian particles. Brownian motion does not conserve momentum unless explicit coupling to a solvent is included; here we consider a so-called ``dry" system for which such coupling is absent or unimportant. (Roughly speaking, this could describe Brownian colloidal particles diffusing close to a wall which acts as a momentum source and sink.) Our order parameter is $\rho$, the particle density. The hydrodynamic equations take the form (with $\J^N = \0$ for the moment)
\begin{eqnarray}
\dot\rho &=& -\del . \J\,, \label{rhocontinuity}\\
\J &=& - M[\rho]\del \mu + \J^N\,,\label{rhocurrent}
\end{eqnarray}
where the collective mobility $M = \rho\beta D[\rho]$ is directly related to the particle diffusivity $D$ which in general depends on where other particles are (and hence on the density field). For simplicity in what follows we make this dependence local: $D[\rho] = D(\rho)$. In \eqref{rhocurrent}, the chemical potential $\mu(\r)$ obeys $\mu(\r) = \delta F/\delta\rho(\r)$ where $F[\rho]$ is a free energy funtional that encodes the interactions between our particles. For a sufficiently soft, pairwise interaction potential $w(r)$ this can be approximated as
\begin{equation}
\beta F[\rho] = \int \rho(\r) (\ln \rho(\r) - 1)\, \d\r + \beta \int \rho(\r_1)\rho(\r_2)w(|\r_1-\r_2|)\, \d\r_1\,\d\r_2 \,,\label{dft}
\end{equation}
where the first term is the entropy of an ideal gas of particles. Note that the form of (\ref{rhocontinuity}) follows from the fact that particles are neither created nor destroyed: $\rho$ is thus a conserved variable and its time derivative must be the negative divergence of some current $\J$. The nonlocal character of the second term in (\ref{dft}) can be approximated in a square gradient theory where one assumes that $\rho$ varies only slowly on the range of the interaction $w(r)$. At this level, we replace (\ref{dft}) with
\begin{equation}
\beta F[\rho] = \int \left( f(\rho) + \frac{\kappa}{2}(\nabla\rho)^2\right) \d\r \,.\label{dft2}
\end{equation}

Adding thermal noise to this model therefore cannot change (\ref{rhocontinuity}), but FDT requires us to choose in (\ref{rhocurrent}) the noise current
\begin{eqnarray}
\J^N &=& \sqrt{2\rho D(\rho)}\bLambda\,,\label{Jnoise}\\
\langle\Lambda_i(\r,t)\Lambda_j(\r',t')\rangle &=& \delta_{ij}\delta(\r-\r')\delta(t-t')\,.\label{unitwhite}
\end{eqnarray}
The quantity $\bLambda$ obeying (\ref{unitwhite}) denotes unit white gaussian vector noise; we will use this notation consistently below (with obvious generalizations to scalar noise $\Lambda$ {\em etc.}). 
Note that in (\ref{Jnoise}) the noise enters multiplied by a function of the field $\rho$. This is called multiplicative noise. In calculations involving the path weights ${\mathbb P}[\rho(\r,t)]$ this is a major technical nuisance; one of the motivations for our next step is to get rid of it.

\subsection{Model B}\label{modB}
To connect with top-down theories based on an expansion in powers of the order parameter and its gradients we proceed in two steps. First we Taylor expand the free energy functional $F[\rho]$ in weak gradients (done already in (\ref{dft2}) and around a reference density $\rho_0$. Writing $\phi = \rho-\rho_0$ this gives the scalar $\phi^4$ free energy
\begin{equation}
F[\phi] = \int {\mathbb F}(\phi,\nabla\phi)\,\d\r = \int \left(\frac{a}{2}\phi^2+\frac{b}{4}\phi^4 +\frac{\kappa}{2}(\del\phi)^2\right)\d\r,
\label{functional}
\end{equation}
Here $a = a(T)$ while $b$ and $\kappa$ are positive and (for simplicity) independent of temperature. Importantly, we have removed a cubic term by selecting as $\rho_0$ the place where $f(\rho)$ has zero third derivative, that is, the critical density. We can also remove a linear term in $\phi$ because its integral over all space is a conserved quantity and hence contributes an additive constant to the free energy. 

Second, we write for the mobility $M(\phi) = \beta(\phi+\rho_0)D(\phi+\rho_0) \simeq \beta\rho_0 D(\rho_0) = M$, a constant, and neglect corrections of order $\phi$. Though technically valid only in uniform phases or close to a critical point where $\phi$ is small everywhere, this simplification is widely used even when describing systems undergoing strong phase separation, because it renders the noise additive rather than multiplicative. Although a somewhat uncontrolled approximation, there are good reasons to think that this does not change the physics qualitatively~\cite{Bray94}.

With these approximations we arrive at ``Model B'':
\begin{eqnarray}
\dot\phi &=& -\del.\J,\label{ModelB}\\
\J &=& -M\del\mu +\J^N, \label{ModelBJ}\\
\mu&=& \frac{\delta F}{\delta\phi} = a\phi + b\phi^3 -\kappa\delsq \phi, \label{ModelBmu}\\
\J^N &=& \sqrt{2Mk_BT}\bLambda \,.\label{ModelBnoise}
\end{eqnarray}
This is the canonical stochastic field theory for diffusive phase separation of a conserved scalar $\phi$ without momentum conservation (which enters through fluid flow; see below). The nomenclature dates back to~\cite{Hohenberg77}; a possible mnemonic is that ``B is for Brownian''.

\begin{figure}
	\centering
	\includegraphics[width=5cm]{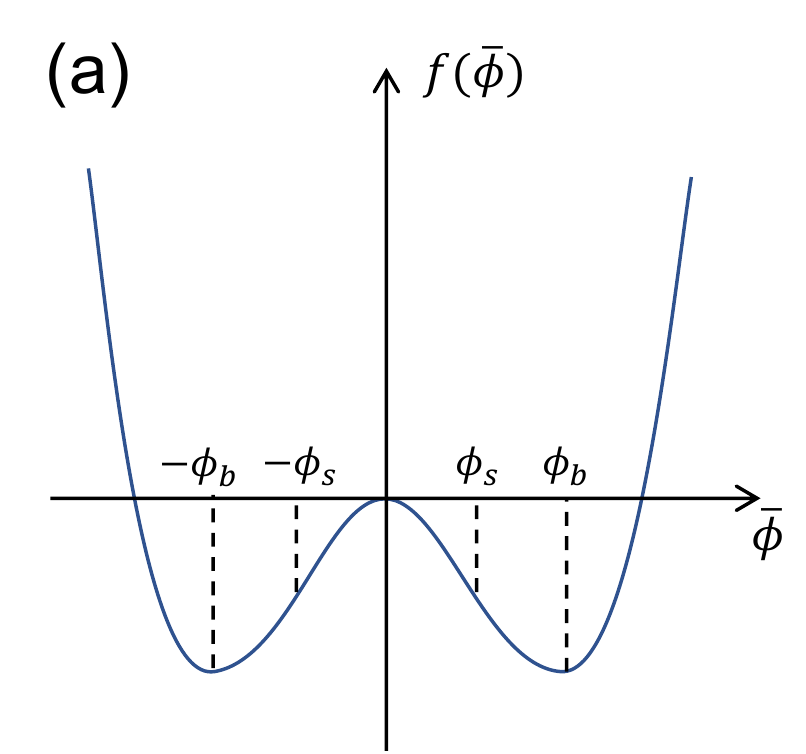}
		\caption[]{Mean-field free energy density for Model B.}
	\label{one}
\end{figure}

The thermodynamics of Model B is straightforward at mean-field level. We consider states of uniform $\phi(\r) = \bar\phi$. For such states
\begin{equation}
\frac{F}{V} = \frac{a}{2}\bar\phi^2+\frac{b}{4}\bar\phi^4 = f(\bar\phi). \label{uniform}
\end{equation} 
For $a>0$ this has a single minimum at $\bar\phi = 0$, with positive curvature everywhere.  The latter means that whatever $\bar\phi$ is chosen, one cannot lower the free energy by introducing a phase separation. On the other hand, for $a<0$, $f$ has negative curvature between the spinodals $\pm\phi_s$ where $\phi_s = (-a/3b)^{1/2}$ (see figure \ref{one}). This signifies local instability. Also, $f$ has two symmetric minima at $\bar\phi = \pm\phi_b$ with $\phi_b = (-a/b)^{1/2}$. For $|\bar\phi|<\phi_b$, $F$ is globally minimized by demixing the uniform state at $\bar\phi$ into two coexisting states at $\phi = \pm\phi_b$. 

Between the binodals, a system of volume $V$ and global composition $\bar\phi$ splits into phases at $\phi = \pm\phi_b$ whose volumes $V_{1,2}$ obey $V_1+V_2 = V$ (conservation of volume) and $(V_1-V_2)\phi_b = V\bar\phi$ (conservation of $\phi$). 
The resulting free energy lies on the horizontal line connecting $f(\pm\phi_b)$. (An additional price must be paid to create an interface between these, but the interfacial area scales as $V^{1-1/d} \ll V$ so that for a large enough system, this price is always worth paying.) More generally, if we had linear or cubic terms in $f(\phi)$, this horizontal line would be replaced by a `common tangent construction' (CTC) whereby a tilted line is tangent to $f(\phi)$ at the coexisting densities $\phi_{1,2}$ and lies below it everywhere else. The CTC arises because both the bulk chemical potential $\mu = df/d\phi$,  and the bulk pressure $P = \phi\mu-f$ (this is a well-known thermodynamic relation) must be equal in two phases at coexistence. Equality of $\mu$ at $\phi_{1,2}$ requires tangents of equal slope; equality of $P$ then requires these to have equal intercept, and hence be a common tangent.
\label{CTCsection}

\subsection{Model H}\label{passiveH}
 
Although set up above for interacting Brownian particles with density $\rho = \phi+\rho_0$, Model B equally describes an interdiffusing incompressible mixture of two species $\alpha$ and $\beta$ in a binary fluid (albeit one without momentum conservation). The order parameter then has a different interpretation, as a composition variable $\phi\sim\rho_\alpha-\rho_\beta$, which again we measure relative to a specific composition $\phi_0$ at which $f(\phi)$ has vanishing third derivative. Since $\phi$ is still a conserved, diffusive field, the form of its equations of motion are retained.

In this setting it is natural to ask how the model generalizes to the `wet' case in which we couple $\phi$ to an incompressible fluid velocity field. (For a one-component system, the fluid flow must inevitably be compressible if the system is to undergo phase separation; that is a more complicated case, which will not concern us here). The resulting generalization (model H; ``H is for hydrodynamics") reads
\begin{eqnarray}
\dot\phi + \v\bdot\del\phi &=& -\del\bdot(-M\del\mu+\J^N),\label{Hphi}\\
\mu(\r) &=&\frac{\delta F}{\delta\phi(\r)} = a\phi+b\phi^3-\kappa\nabla^2\phi,\label{Hmu}\\
\rho(\dot\v+\v\bdot\del\v) &=& \eta\delsq \v-\del P +\del\bdot\bSigma^\phi +\del\bdot\bSigma^N\,.\label{HNSE}
\end{eqnarray}
In the equation of motion for $\phi$, (\ref{Hphi}), we have simply replaced the time derivative by one co-moving with the fluid on the left hand side. (With $\del\bdot\v = 0$ as required  can be alternatively viewed as the divergence of an advective current $\v\phi$.) In the NSE we have, alongside the noise stress $\bSigma^N$ specified in (\ref{Hvnoise}), an additional term
\begin{equation}
\del\bdot\bSigma^\phi = -\phi\nabla\mu\,. \label{phigradmu}
\end{equation}
Th stress $\bSigma^\phi$ stems from the order parameter via the free energy $F[\phi]$ and hence, ultimately, stems from interparticle forces. It tells the fluid how to move so as to reduce the free energy stored in `elastic' distortions of the order parameter field. (Technically the NSE with this term is no longer the Navier-Stokes equation but a Cauchy equation. We ignore this distinction in all that follows.)

To derive the equality in (\ref{phigradmu}), consider
a small incompressible displacement field $\u$: that is $\r\to\r+\u(\r)$ with $\del\bdot\u = 0$. Advection of the $\phi$ field by this displacement induces the change $\phi(\r) \to \phi(\r-\u)$.  To linear order this gives the increment $\delta\phi  = -\u\bdot\del\phi$ from which we find the free energy increment as 
\begin{equation}
\delta F = \int\frac{\delta F}{\delta \phi}\delta\phi\,\d\r=-\int \mu\u\bdot\del\phi\, \d\r = \int (\phi\del\mu)\bdot\u\,\d\r,\label{stress1}
\end{equation}
where the final form follows by partial integration and incompressibility. (We consider periodic boundary conditions without loss of generality; this eliminates the boundary term.) This result can be compared with the free energy increment caused by a strain tensor $\bepsilon = \del\u$
\begin{equation}
\delta F = \int \bSigma^\phi\bcol(\del\u)\,\d\r = - \int \del\bdot\bSigma^\phi\bdot\u\,\d\r ,\label{stress2}
\end{equation}
where the second form again follows by partial integration. Comparison of (\ref{stress1},\ref{stress2}) establishes (\ref{phigradmu}).
This form, which is all we need to know about the stress tensor for the purposes of solving the NSE, could also have been derived from the following expression for the stress tensor itself:
\begin{equation}
\Sigma_{ij} = -P^\phi\delta_{ij} -\kappa(\partial_i\phi)(\partial_j\phi).\label{stress3}
\end{equation}
Here $P^\phi= \phi\mu - {\mathbb F}$ (where ${\mathbb F}$ obeys (\ref{functional})) is the order parameter contribution to the (local) pressure, which in the binary fluid setting is usually called the osmotic pressure.

\subsection{Nonconserved order parameters}

Nonspherical molecules may need additional order parameters to describe their orientations. Since each molecule can rotate locally, these are not conserved. 

For example we might have a vector field $\p$, describing the mesoscopic mean orientation of rodlike molecules:
\begin{equation}
\p(\r) = \langle \nuhat\rangle_{\rm meso},\label{pdef}
\end{equation}
with  $\nuhat$ a unit vector along the axis of a single molecule. A material of nonzero $\p$ is called a polar liquid crystal. This order parameter makes sense only for molecules that have one end different from the other. Even in that case $\p$ vanishes when molecules are oriented but not aligned, in the sense that they point preferentially along some axis but are equally likely to point up that axis as down it. To describe cases with orientation but not alignment (in the sense just defined), we need a second rank tensor
 \begin{equation}
\Q(\r) = \langle \nuhat\nuhat\rangle_{\rm meso}-\I/d,
\label{Qdef}
\end{equation}
where $\nuhat\nuhat$ is a dyadic product (and independent of which way the unit vector points along the molecule); $\I$ is the unit tensor, and $d$ is the dimension of space. The resulting tensor is traceless by construction and therefore vanishes if the rods are isotropically distributed. A fluid in which $\Q$ is finite but $\p$ is zero is called a nematic liquid crystal.

The equation of motions in each case are quite simple for dry systems but significantly more complicated when coupled to fluid flow. Here we address both cases for the case of a polar liquid crystal. Supposing the rotational motion of $\p$ to be overdamped, we can write in the dry case
\begin{equation}
\dot\p = -\Gamma\frac{\delta F}{\delta \p} + \sqrt{2k_BT\Gamma} \bLambda\,. \label{dryp}
\end{equation}
The first term is not of the form of a divergence of a current because there is no conservation law on $\p$. It is instead a simple relaxation term that aims to reduce the free energy, with a rotational mobility $\Gamma$. This describes a rotational friction and hence is accompanied necessarily by the noise term shown. (Below, this FDT relation will be made explicit for a closely related model.) 

The quantity $\delta F/\delta\p$ is often called the `molecular field'.
By choosing the following form (which is not the most general, replacing several independent square-gradient terms with a single combination)
\begin{equation}
F[\p] =  \int \left(\frac{a}{2}|\p|^2+\frac{b}{4}|\p|^4 +\frac{\kappa}{2}\del\p\bcol\del\p\right)\d\r,
\label{pfunctional}
\end{equation}
we can set $a<0$ to describe a state of spontaneous polarization. This polarization, if uniform, can point in any direction without changing $F$ -- so we have a state of spontaneously broken rotational symmetry. This is a continuous symmetry and therefore qualitatively different from the discrete symmetry of the $\phi^4$ scalar theory. Note however that the scalar version of (\ref{dryp}), namely
\begin{equation}
\dot\phi = -\Gamma\frac{\delta F}{\delta \phi} + \sqrt{2k_BT\Gamma} \Lambda\,. \label{A}
\end{equation}
describes the non-conserved dynamics of (say) a scalar magnetization in a ferromagnetic system (where spins can flip rather than diffuse). This is called `Model A' in the classification of 
\cite{Hohenberg77}. We return to it later (in Section \ref{ch:four}).

We turn now to the case of a `wet' polar liquid crystal. For this we can write
\begin{eqnarray}
\frac{D\p} {Dt} &=& -\Gamma\frac{\delta F}{\delta \p} + \sqrt{2k_BT\Gamma} \bLambda,\label{wetp}\\
\rho(\dot\v+\v\bdot\del\v) &=& \eta\delsq \v-\del P +\del\bdot\bSigma^\p +\del\bdot\bSigma^N\,.\label{pNSE}
\end{eqnarray}
Here we need to specify both the proper form of the comoving derivative, $D\p/Dt$, and the order parameter stress $\bSigma^\p$. Within $D\p/Dt$ one expects a term like $\v\bdot\del\p$ which is trilinear. For a scalar field this was the only trilinear term, but for a vector, there are others. These describe the rules whereby the vector is rotated, rather than translated, in the presence of a flow field. The general form is found to be
\begin{equation}
\frac{D\p}{Dt} = \dot\p + \v\bdot\del\p + \bOmega\bdot\p -\xi\D\bdot\p\,,\label{advp}
\end{equation}
where $\D$ and $\bOmega$ are respectively the symmetric and antisymmetric parts of the velocity gradient tensor $\del \v$. 
In (\ref{advp}), the unity prefactor of the $\v\bdot\del\p$ term is required by Galilean invarance, while that of the $\bOmega\bdot\p$ is required by the fact that if, in a system of uniform $\p$, the entire sample is subject to a rigid-body rotation, $\p$ co-rotates with the rigid body. In contrast to these two terms, $\xi$, sometimes called a slip parameter and entering here with a conventional minus sign, depends on molecular physics and describes how particles rotate in a shear flow. 

Having specified (via $\xi$) how $\p$ is advected when the sample changes shape, it is possible to re-work the arguments leading to (\ref{stress1},\ref{stress2}) for the stress of a scalar field to give the corresponding expressions for a vector $\p$. Details can be found in~\cite{CatesTjhungJFM}. The result is, in terms of the molecular field $\h = \delta F/\delta \p$, 
\begin{eqnarray}
\nabla_i\Sigma^\p_{ij} = &-&p_k\nabla_jh_k\\
&+& \nabla_i (p_ih_j-p_jh_i)/2\\
&+& \xi\nabla_i (p_ih_j+p_jh_i)/2\,,
\end{eqnarray}
whose three contributions respectively stem from the three advective terms on the right hand side in (\ref{advp}). Note the similarity in form between the $-p_k\nabla_jh_k$ and $-\phi\del\mu$ for a scalar field. Note also that the `flow rule' encoded in $\xi$ inevitably also enters the elastic stress since it specifies how to update the order parameter when the sample changes shape.

\section{Active versus passive field theories}\label{actpass}\label{ch:two}
So far, we have given example of stochastic field theories for scalar and vector order parameters $\Psi$ describing various types of system (wet/dry, conserved/nonconserved). All the results given above are part of the classical literature for field theories of soft matter in the absence of activity. We now need to see what makes active systems different.

The key difference is that passive field theories are constrained in form by the time-reversal symmetry (TRS) of the underlying microscopic dynamics. These constraints are lifted for active materials which dissipate a continuous supply of energy by converting  it into motion at the particle scale. The consequences of TRS (which are all interlinked) include the following:

\noindent(i) The existence of a free energy functional $F[\Psi(\r)]$.

\noindent(ii) The existence and uniqueness of the Boltzmann distribution $P_0\sim \exp[-\beta F]$ in steady state.

\noindent(iii) The principle of detailed balance (PDB). This states that 
\begin{equation}
P_0[\Psi_1]{\mathbb P}[\Psi_1\to\Psi_2; t] = P_0[\Psi_2]{\mathbb P}[\Psi_2\to\Psi_1; t]\,, 
\label{PDB}\end{equation}
where ${\mathbb P}$ is a path probability. In words, this means that in steady state, the probability of seeing the system transition from state $\Psi_1$ to $\Psi_2$ (via any chosen path or set of paths) is exactly the probability of seeing the reverse process. Put differently, any steady-state movie looks statistically the same when run in reverse.

\noindent(iv) There are no circulating currents in steady state, either in real space of in the space of configurations. (This follows directly from (iii).) 

\noindent(v) The Fluctuation-Dissipation Theorem (FDT). This name is used for various different results in statistical physics, but here we use it to mean the fact that in a stochastic field theory for $\Psi(\r,t)$, the noise terms causing fluctuations are fixed by the form of the dissipative coefficients in the deterministic part of the equation. 

It is a useful excursion to prove the FDT explicitly, at least for the simplest case of Model A. The generalization to other stochastic field theories for passive matter (Model B, Model H, dry or wet liquid crystals) is reasonably straightforward. Rewriting the equation of motion for Model A as
\begin{equation}
\dot\phi = -\Gamma\frac{\delta F}{\delta \phi} + \nu \Lambda\,, \label{Anu}
\end{equation}
our job is to prove $\nu = \sqrt{2k_BT\Gamma}$. We first rewrite (\ref{Anu}) with the noise $\Lambda$ as the subject:
\begin{equation}
\Lambda = \frac{1}{\nu}\left(\dot\phi+\Gamma\frac{\delta F}{\delta\phi}\right)\,.\label{Lambdasubject}
\end{equation}
We now observe that for unit white gaussian noise $\Lambda$, 
\begin{equation}
{\mathbb P}[\Lambda(\r,t)] = {\cal N}_\lambda \exp\left[-\frac{1}{2}\int \Lambda^2\d\r\d t\right], \label{whiteweight}
\end{equation}
where ${\cal N}_\lambda$ is a normalization. Substituting (\ref{Lambdasubject}) into (\ref{whiteweight}) gives the path weight for a trajectory $\phi(\r,t)$ in the forward dynamics between an initial state and time (1) and a final state and time (2)
\begin{equation}
{\mathbb P}_F[\phi(\r,t)] = {\cal N}_\phi \exp\left[-\frac{1}{2\nu^2}\int_{(1)}^{(2)} \left(\dot\phi+\Gamma\frac{\delta F}{\delta \phi}\right)^2\d\r\d t\right], \label{forwardweight}
\end{equation}
where $N_\phi$ is another normalizer (absorbing a Jacobian factor). On the other hand, the weight for the reverse path is found by setting $\dot\phi \to - \dot\phi$ in this expression, because the time derivative is negated for the reversed path while the field variables are not:
\begin{equation}
{\mathbb P}_R[\phi(\r,t)] = {\cal N}_\phi \exp\left[-\frac{1}{2\nu^2}\int_{(1)}^{(2)} \left(-\dot\phi+\Gamma\frac{\delta F}{\delta \phi}\right)^2\d\r\d t\right], \label{backwardweight}
\end{equation}
We can therefore construct the ratio of the forward and backward path weights that carry us from state (1) to state (2) or vice-versa as
\begin{equation}
\frac{{\mathbb P}_F}{{\mathbb P}_B} = \exp\left[-\frac{\Gamma}{2\nu^2}\,4\int_{(1)}^{(2)} 
\dot\phi\frac{\delta F}{\delta \phi}
\d\r\d t\right]\,. \label{ratioweight}
\end{equation}
This stems from the two cross terms arising when we expand the squared terms in (\ref{forwardweight},\ref{backwardweight}), with all other terms cancelling between the forward and backward paths.

Now the spatial integral in (\ref{ratioweight}) gives us simply the time derivative of the free energy; performing also the time integral we have
\begin{equation}
\frac{{\mathbb P}_F}{{\mathbb P}_B} = \exp\left[-\frac{2\Gamma}{\nu^2}\left(F[\phi_2]-F[\phi_1]\right)\right]\,. \label{ratioweight2}
\end{equation}
The principle of detailed balance (\ref{PDB}) then demands that this ratio equates to the ratio of Boltmzann factors for final and initial states
\begin{equation}
\frac{P_0[\phi_2]}{P_0[\phi_1]} =  \exp\left[-\beta\left(F[\phi_2]-F[\phi_1]\right)\right]\,.
\end{equation}
It follows that  $\nu = \sqrt{2k_BT\Gamma}$ which is the required result.

\subsection{Construction of active field theories}
Active field theories involve, as equations of motion, stochastic PDEs for suitable order parameter fields $\Psi(\r,t)$ (such as $\rho, \v,\phi, \p,\Q$) that are unconstrained by the requirements (i-v) above that all stem from microscopic time-reversal symmetry.

There are two basic routes to the construction of such theories, as indeed there are in the passive case. Either we can construct them bottom-up by explicit coarse-graining of a microscopic (e.g., particle-level) model, or top-down by writing down all terms allowed by symmetry and conservation laws to given order, and then adding noise. The bottom-up route is unambiguous, and leads to definite parameter values in the final theory. On the other hand it is hard work, and there is no guarantee that the chosen micro-model generates all the important terms for a general continuum theory. The top down route is relatively easy, and fairly systematic; however it can generate numerous unknown parameters and there is no FDT to guide the choice of noise terms.

Once such a field theory is constructed it can usually be viewed as
\begin{equation}
\dot\psi = [{\rm effectively~passive~terms}] + ({\rm activity~parameters})\times({\rm explicit~TRS~violations}).\label{decomp}
\end{equation}
This decomposition is useful when the resulting passive terms corresponds to a model that is already understood. Note however the the ``effectively passive'' sector may include {\em implicitly} active mechanisms. An example is motility-induced phase separation, which we consider in depth later on. But this raises a question: when do the {\em explicit} TRS breaking terms control the physics at large scales? When they do, the system is irreducibly active in its behaviour. When they don't, the system is quasi-passive in some sense.

In Section \ref{ch:three} we will carry out both the bottom-up and top-down approaches for the case of a dry active scalar. First, we survey various other cases, top-down only.

\subsection{Dry flock}
We outline here a version of the Toner-Tu model for which the decomposition (\ref{decomp}) is apparent; it differs somewhat from the original~\cite{TonerTu} but closely follows the presentation in~\cite{Marchetti13}. This model we select has the following structure:
\begin{eqnarray}
\dot\rho &=& -\del\bdot\J\,,\\
\J &=& \J^a+\J^p\,,\label{TTflux}\\
J^a &=& v_0\rho\p\,,\\
J^p &=& -M\del\frac{\delta F}{\delta\rho} + \sqrt{2k_BTM}\bLambda\,,\\
\dot\p +\lambda_1\p\bdot\nabla\p &=& - \Gamma\frac{\delta F}{\delta\p} + \sqrt{2k_BT\Gamma}\bLambda\,,\label{TTpdot}
\end{eqnarray}
with two independent activity parameters, $\lambda_1$ and $v_0$; the {\em effective} free energy is
\begin{equation} 
F = \int\left(\frac{a}{2}|\p|^2 + \frac{b}{4}|\p|^4 +\frac{\kappa}{2}\del\p\bcol\del\p - \bar w(\del\bdot\p)\rho +\frac{\lambda}{2\Gamma}|\p|^2\del\bdot\p\right)\,\d\r\,.\label{TTF}
\end{equation}

The particle flux in (\ref{TTflux}) has an active contribution $\J^a$ whereby particles move with speed $v_0$ along their molecular axes so that the mass flux is proportional to the local polarization $\p$. As written, it also has a passive contribution $\J^p$; the original Toner-Tu treatment set this to zero (effectively choosing $M = 0$) but retaining $M$ allows the passive limit of the resulting model to be taken in a nonsingular fashion. 
Equation \ref{TTpdot} can be written as
\begin{equation}
\dot\p +\lambda_1\p\bdot\nabla\p + \lambda_2\p\del\bdot\p + \frac{\lambda_3}{2}\del|\p|^2 =\\
-\Gamma[a\p+b|\p|^2\p -\kappa\nabla^2\p +\bar w\del\rho]+\sqrt{2k_BT\Gamma}\bLambda\label{TTpdotfull}
\end{equation}
in which we make the specific choice, $\lambda_3=-\lambda_2 = \lambda$. This choice is motivated by microscopics (see~\cite{Marchetti13}) and implied by the effective free energy (\ref{TTF}); the fully general case has $\lambda_{2,3}$  independent, effectively introducing a third activity parameter. Note how the effective free energy (\ref{TTF}) includes a term in $\lambda$ which is of passive form but of active origin -- a `quasi-passive' term.

In (\ref{TTpdotfull}), $\lambda_1$ describes advection of particles by propulsion along their axes; this resembles the Navier-Stokes nonlinearity $\v\bdot\del\v$ but $\lambda_1$ need not be unity (essentially because there is no Galilean invariance in this dry system~\cite{TonerTu}). The terms in $a,b,\kappa$ are standard contributions to the molecular field; negative $a$ creates a state of spontaneous polarization. The $\bar w$ term is also standard in passive liquid crystals; it creates a tendency for the polarization to align with or against density gradients. The physics behind the terms in $\lambda_{2,3}$ is considered in the lectures by John Toner in this volume, where the surprising collective behaviour of the model is also discussed. This includes the ability to create true long range order in two dimensions, an outcome forbidden in equilibrium for systems with a spontaneously broken rotational symmetry. Other features include extreme density fluctuations (also known as giant number fluctuations) and a transition from an isotropic state to one of moving stripes~\cite{Marchetti13}. The latter state represents a form of microphase separation even though there is nothing in the effective free energy (\ref{TTF}) that would cause this in equilibrium. We shall see a similar feature of `unexpected microphase separation' emerging from a scalar model in Section \ref{ch:six} below.

\subsection{Dry active nematic}
The nematic order parameter $\Q$ was defined in (\ref{Qdef}) to describe an oriented state of headless rods (or of headed rods whose vectors point with equal probability up or down the major axis of $\Q$). We here construct the active theory top-down as a passive model to which leading-order active terms are added. We write at deterministic hydrodynamic level
\begin{eqnarray}
\dot\Q &=& - \Gamma\H\,,\\
H_{ij} &=&\frac{\delta F}{\delta Q_{ij}} - \left(\Tr \frac{\delta F}{\delta Q_{ij}}\right)\frac{\delta_{ij}}{d}\,.
\end{eqnarray}
where the second term in the molecular field $\H$ ensures that $\Q$ remains traceless under time evolution. With a similar simplification as used previously for the case of a polar liquid crystal (replacing three separate elastic constants with a single combination) we may choose for the effective free energy
\begin{eqnarray}
F[\Q,\rho] &=& \int\left(a\Tr(\Q^2) + b_1(\Tr(\Q^2))^2 + b_2\Tr(\Q^4)
+ c\Tr(\Q^3) + \frac{\kappa}{2}(\nabla_iQ_{jk})(\nabla_iQ_{jk})\right)\d\r \nonumber\\
&+& \int\left(W\Q\bcol\del\del\rho + \frac{A}{2}\rho^2\right)\,\d\r\,.\label{QF}
\end{eqnarray}
The terms in the top line of (\ref{QF}) are those of a standard passive nematic. Those in the second line are also present in general for a compressible passive nematic, with $W$ the lowest-order coupling constant between the nematic order parameter $\Q$ and the density $\rho$. The term in $A$ is a harmonic restoring force that opposes density fluctuations; so long as this coefficient is sufficiently positive we do not expect phase separation in this model. Note that to leading order, the dynamics of $\Q$ written above is entirely governed by quasi-passive terms.

\begin{figure}
	\centering
	\includegraphics[width=4cm]{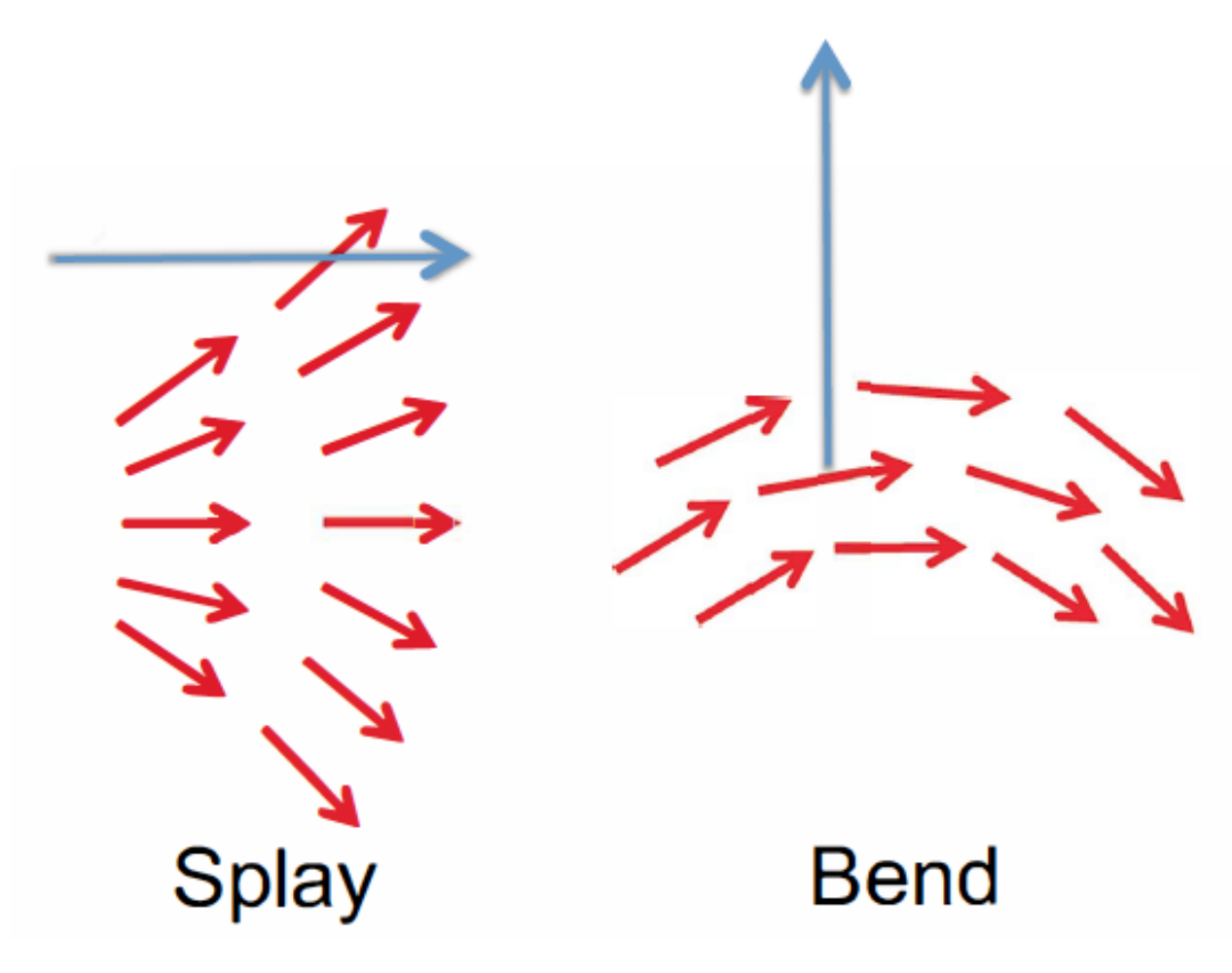}\hspace{1cm}
	\includegraphics[width=5cm]{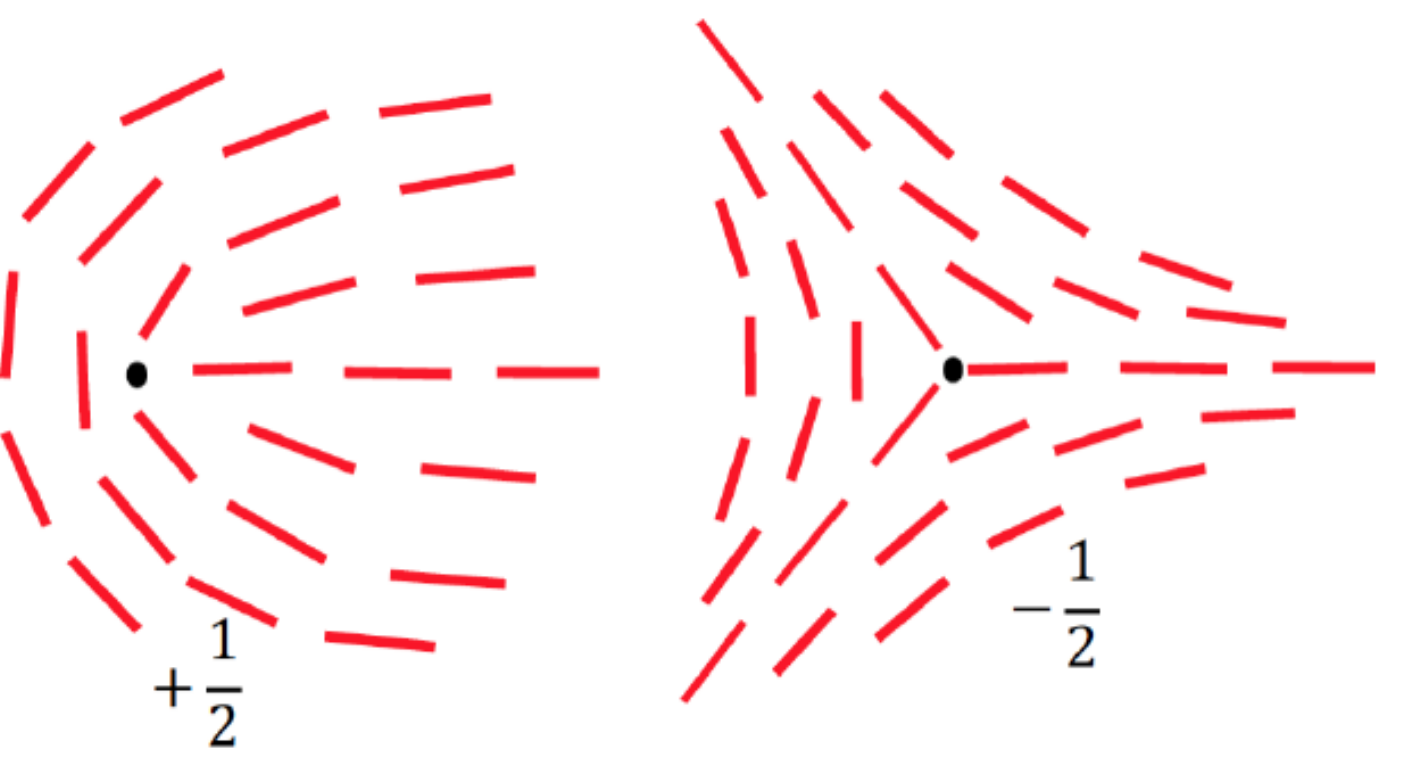}
	\caption[]{Left two frames: Splay and bend deformations, with the large arrow (or its reverse, depending on the sign of $\zeta$) indicating the mass-flux induced in a dry active nematic by spatial variations in the order parameter field. Right two frames: topological defects of charge $\pm 1/2$ in a nematic system. The charge refers to the fact that on rotating one full turn about the defect core (black dot), the orientation of the molecules rotates by a half turn, either in the same direction ($+$) or the opposite one ($-$). The active mass flux shown in the left frames causes the $-1/2$ defect to move along or against the direction of its `arrow', whereas the threefold symmetry of the $+1/2$ defect means that there is no net active motion.}
	\label{splaybend}
\end{figure}

Indeed, the leading-order violation of TRS in this hydrodynamic-level model lies not in the equation of motion for $\Q$ but in that for the density, which reads as usual $\dot\rho = -\del\bdot \J$ with  
\begin{equation}
\J =  - M\del\frac{\delta F}{\delta\rho} + \zeta \del\bdot\Q\,,\label{DryQJ}
\end{equation}
where the first term on the right is the usual passive diffusive current and the term in $\zeta$ is the lowest-order active current contribution, within a simultaneous Taylor expansion in $\rho, \Q$ and $\nabla$. Depending on the sign of $\zeta$, this term creates a mass flux either along or against a direction in which $\Q$ has a gradient. This axis is shown in Fig.~\ref{splaybend} for the cases where molecular orientation is subject to a splay deformation or bend deformation. In an equilibrium system, such configurations generate a stored elastic energy but not a mass flux (so the $\zeta$ term cannot derive from a free energy). It should be obvious that a steady state movie, when such a flux is present, does not look the same running backwards and therefore this form of mass flux indeed violates TRS. 

Among interesting consequences of the activity are (a) giant number fluctuations -- as also seen in the Toner-Tu model and its relatives, discussed previously --  and (b) self-propulsion of topological defects (see Fig.~\ref{splaybend}). This type of behaviour is seen experimentally in dry active nematics such as vibrated granular rods~\cite{Marchetti13} and also is seen in wet systems where the active mass-flux gets replaced by an active mechanical stress that leads to broadly similar consequences~\cite{Giomi13,Sanchez12}.

\subsection{Wet active liquid crystals and further variants}
At this point it would be natural to give the hydrodynamic-level equations of motion for wet active liquid crystals (both polar and nematic). However these would lead us further in the direction of generating more and more complicated equations. Suffice to say that, for the polar case whose passive limit was already considered in (\ref{wetp},\ref{advp}), additional self-advective terms in the $\dot\p$ equation arise from activity, resembling those in the dry flock counterpart (\ref{TTpdotfull}). In addition, there is an active stress in the Navier Stokes sector, which must take the form of the divergence of a stress, as usual. The active stress takes the form
\begin{equation}
\Sigma_{ij}^A = -\zeta p_i p_j + \chi(\nabla_ip_j+\nabla_jp_i)\,,\label{swimstressp}
\end{equation}
where the minus sign before $\zeta$ is conventional and the second term is subdominant to the first in many situations. (This $\zeta$ has no direct connection with the one in (\ref{DryQJ}).) The $\zeta$-stress describes the mechanical forcing on the fluid caused by a self-propelled particle that is extensile ($\zeta> 0$: fluid is pulled in around the equator of the particle and expelled along its axis of motion) or contractile ($\zeta<0$, with the reverse local fluid flow). A similar stress term appears for wet nematics, now taking the form $-\zeta\Q$. Further discussion of such cases can be found in~\cite{Marchetti13,CatesTjhungJFM}. The first of these references also considers cases where both polar and nematic order are simultaneously present; where a wet $\p$ or $\Q$ is coupled to a scalar (a density or composition field); or where the NSE is replaced by a viscoelastic flow model. These are very interesting areas but in the rest of this article we will go back to basics and try to understand what we can about the simplest type of active field theory when only a single scalar order parameter is present.

Before doing that, we mention one distinct way in which TRS can be broken in systems where there is more than one order parameter (such as wet active liquid crystals which have a fluid velocity $\v$ alongside $\p$ or $\Q$). Namely, one could imagine a system in which all the equations of motion have quasi-passive form, but the free energy functionals that generate the passive terms are different from one equation to another. An instance of this will arise later on in the case of an active scalar field coupled to a fluid velocity (``Active Model H" in Section \ref{ch:five}).

\section{From scalar active particles to scalar field theory}\label{ch:three}
We have mentioned a number of physical effects arising in the active models surveyed above. On of these is giant number fluctuations whose most extreme form (fluctuations in density of order the mean) corresponds to `unexpected phase separation'. Here the word `unexpected' means `not arising in passive counterpart systems of comparable simplicity'. A second effect is  `unexpected microphase separation', in which rather than separating into macroscopic domains of high and low density the system maintains a pattern of contrasting densities on a finite length scale (such as a phase of stripes, which is seen in dry flocks~\cite{Marchetti13}.
A third is the existence of circulating currents in steady state -- a frequent occurrence in say 2D layers of swimming bacteria where, confined in a closed container, one can see a clockwise or counterclockwise current that persists indefinitely~\cite{Goldsteinflux}.  A general question is then: {\em how much of the generic physics of active systems is already present in the simplest active field theories, governed by a single scalar $\rho$ or $\phi$?} To answer this question, we consider in this rest of these lectures the construction and behaviour of such theories.

\subsection{Explicit coarse-graining of particle dynamics}
We start from a model of self-propelled spheres (or discs in 2D). The $\alpha$th particle has a position vector $\r_\alpha$, and an orientational vector $\u_\alpha$ which denotes the direction of its self-propulsion. We assume there are no alignment interactions (this neglects near-field hydrodynamics, for example~\cite{FieldingNF}), which means that the orientational dynamics of each particle is independent of all others. Accordingly there can be no mean polarization or nematic order in bulk phases: $\p =\Q = \0$. The only slow variable is therefore the density $\rho$, at least for the dry case which we consider here.

We start with a single particle and suppose it to have a propulsion speed $v_0(\r)$ along $\u$, which undergoes rotational Brownian motion with angular diffusivity $D_r$. The probability density $\psi(\r,\u,t)$ ({\em not} to be confused  with an order parameter, despite this notation) for finding the particle at $\r,\u,t$ obeys
\begin{equation}
\dot\psi = -\del\bdot(v_0\u\psi) + \partial_\u\bdot(D_r\partial_\u\psi)\label{ABP1}
\end{equation}
where the first term is propulsion and the second is rotational diffusion. Here $\del$ is a spatial gradient and
 $\partial_\u = \hat{\boldsymbol{\theta}}\partial_\theta + (\sin \theta)^{-1}\hat{\boldsymbol{\phi}}\partial_\phi$ is the gradient operator on a unit sphere 
(with the second term absent in 2D). 

One way to write the spherical harmonic expansion of $\psi$ is as
\begin{equation}
\psi(\r,\u,t) = \tilde\rho + \tilde p_iu_i +\tilde{Q}_{ij}(u_iu_j-\delta_{ij}/d) + ...
\label{psiexpand}\end{equation}
where higher order terms are neglected. (The unimportance of these terms is shown in~\cite{Cates_Tailleur_EPL}, whose presentation we simplify here.)  Here $\tilde\rho = S_d^{-1}\int\psi\d\u$, {\em etc.}, with $S_d$ the surface area of the unit sphere in $d$ dimensions. Integrating first $\dot\psi$ in (\ref{psiexpand}) over the unit sphere, and then $\dot\psi\u$ likewise, gives us two equations:
\begin{eqnarray}
\dot{\tilde \rho} &=& -\frac{1}{d}\del\bdot(\v_0\tilde\p) \,,\label{moment1}\\
\dot{\tilde \p} &=& -\nabla(v_0\tilde\rho) - D_r(d-1)\tilde\p + ...\label{moment2}
\end{eqnarray}
Of these two variables, $\tilde\rho$ is slow and conserved, while $\tilde\p$ is neither. We can therefore make an adiabatic approximation, $\tilde \p = -\nabla(v_0\tilde\rho)\tau$ where the angular relaxation time obeys $\tau^{-1} = D_r(d-1)$. (This result is easily extended to the case where angular relaxation is not diffusive but discontinuous, for example by a `run-and-tumble' dynamics; see~\cite{Cates_Tailleur_EPL}.) Accordingly
we may write \begin{eqnarray}
\dot{\tilde \rho} &=& -\del\bdot\J_1 \,,\label{current1}\\
\J_1 &=& -\v_0\tau \nabla(v_0\tilde\rho) = \V\tilde\rho -D\nabla\tilde\rho\,.\label{current2}
\end{eqnarray}
Here the subscript $1$ reminds us that this is a one-particle probability current, not a particle current. In the final equality we introduce drift and diffusivity variables
\begin{equation}
\V(\r) = -\frac{\tau}{d}v_0\nabla v_0\;\;\;;\;\;\;D(\r) = \frac{\tau}{d}v_0^2\,, \label{drift1}
\end{equation}
each of which depends on position via $v_0(\r)$. 

The one-particle probability current $\J_1$ obeying (\ref{current2}) represents a coarse-grained description of the particle's behaviour, because we have neglected higher terms in the spherical harmonics of $\psi$. Having arrived at this description, we notice that it is {\em exactly the same} as a passive Brownian particle of diffusivity $D(\r)$ subject to the drift $\V(\r)$. This is in turn described by an It\^{o}-Langevin equation
\begin{equation}
\dot\r = \A(\r) + C(\r)\bLambda \;\;;\;\; \A \equiv \V +\nabla D \;\;;\;\;C \equiv \sqrt{2D}\label{ILE}
\end{equation}
where the It\^{o} interpretation requires that in the multiplicative noise $C$ is evaluated at the start of each time step (in a discrete representation). Having chosen this interpretation, we can easily now generalize to the case of $N$ interacting particles whose speeds $v_0$ depend on position, not through a pre-assigned function as in (\ref{drift1}) but via information about where all the other particles are. For simplicity and definiteness, we introduce a so-called `quorum-sensing' interaction whereby 
\begin{equation}
v_0(\r) = v(\r,[\rho])\label{QS}
\end{equation}
where 
\begin{equation}
\rho(\r) = \sum_\alpha\rho^\alpha(\r)\;\;;\;\; \rho_\alpha \equiv\delta(\r-\r^\alpha)\,,\label{empirical}
\end{equation}
with $\alpha$ a particle label. The `empirical density' $\rho$ is highly singular and contains full information about particle positions, but soon we will coarse-grain further and view it as a smoothly varying order parameter, the particle density.

For our quorum-sensing particles we now have
\begin{equation}
\dot\r^\alpha = \A(\r^\alpha,[\rho]) + C(\r^\alpha,[\rho]))\bLambda \,.
\end{equation}
A theorem (It\^{o}'s theorem) then states that for any function $f(\r)$ the corresponding It\^{o}-Langevin equation is
\begin{equation}
\dot f(\r) = (\A + C\bLambda)\bdot\frac{\partial f}{\partial\r} + \frac{C^2}{2}\frac{\partial^2f}{\partial\r}\,,
\end{equation}
which can be rewritten for each $\alpha$ as
\begin{equation}
\dot f(\r^\alpha) = \int\rho^\alpha(\r,t)\left[(\A + C\bLambda^\alpha)\bdot\nabla f(\r) +
D\nabla^2f(\r)\right]\d\r\,.
\end{equation}
Following an imaginative procedure introduced by Dean~\cite{Dean} we now make the choice $f(\r^\alpha) = \rho^\alpha$, integrate by parts, and note that $\dot f(\r^\alpha) = \int f(\r)\dot\rho^\alpha(\r)\,\d\r$ to obtain
\begin{equation}
\dot\rho(\r^\alpha) = -\del\bdot(\A\rho^\alpha) + \nabla^2(D\rho^\alpha) - \del\bdot(\bLambda^\alpha C\rho^\alpha)\,.
\end{equation}
Again following Dean, we sum on $\alpha$ and read the resulting equation as the Langevin dynamics for a smooth order parameter field $\rho$:
\begin{eqnarray}
\dot\rho &=& -\del\bdot\J\,,\\
\J &=& \A\rho -\nabla(D\rho) - \sum_\alpha\bLambda^\alpha C\rho^\alpha\,,\\
\Rightarrow \J &=& \V\rho - D\nabla\rho +\sqrt{2D\rho}\bLambda\,. \label{finalDean}
\end{eqnarray}
In the last line we have substituted the definition of $\A$ from (\ref{ILE}), and in each mesoscopic region replaced a sum of independent white noises with a single noise of the same variance. 

We have arrived in (\ref{finalDean}) at a Langevin equation for the collective density of precisely the same form as a set of passive Brownian particles with interacting forces and spatially varying diffusivity governed by $\V, D$ obeying (\ref{drift1},\ref{QS}). Specifically the force acting on a particle at $\r$ is
\begin{equation}
\f(\r,[\rho]) = k_BT\frac{\V}{D} = -k_BT \nabla \ln v(\r,[\rho])\,,
\end{equation}
so that we can write the current in terms of an effective chemical potential
\begin{eqnarray}
\J &=& -\rho D \beta\nabla\mu + \sqrt{2D\rho}\bLambda\,,\\
\beta\mu &=& \ln \rho +\ln v[\rho]\,.\label{chempot}
\end{eqnarray}
In (\ref{chempot}), the first term describes an ideal gas, and the second comes from the combination of self-propulsion and quorum sensing. Note that in $\J$ all factors of $k_BT$ cancel out, as they clearly must for a model in which temperature did not originally enter the microscopic dynamics defined via (\ref{ABP1}).

The final answer (\ref{finalDean}) was possibly a guessable result, but we have given a full derivation because this is a rare case where coarse-graining can be completed without ad-hoc approximations. (Note however that a pure mathematician would regard the Dean procedure as ad-hoc, or worse!) The resulting equation should apply at long times whenever the length-scale $v_0\tau$ is small compared to any other relevant length. Such lengths include the interparticle distance, the length set by the range of the quorum sensing interaction, and the length-scale over which $\rho$ varies appreciably. If in practice these stringent conditions are not met, then we are back in the domain of ad-hoc  (though hopefully good) approximation. In particular, the same theory can be used approximately to describe active Brownian particles interacting via collisions rather than quorum sensing. Collisions also cause the mean speed in the direction of $\u$ to be a decreasing function of density, but their range is short compared to $v_0\tau$ so the formal requirements of the coarse-graining procedure made above are not met.

We see from (\ref{chempot}) that in steady state, $\rho v$ will be constant. That is, particles accumulate where they move slowly. But also, for typical types of quorum sensing (or collisions) $v$ decreases with $\rho$. This creates the positive feedback which can cause motility-induced phase separation (MIPS)~\cite{Cates15}.

\subsection{Mean-field theory for MIPS}\label{MFMIPS}
We consider first uniform phases in steady state. Here $v[\rho] = v(\rho)$ is constant within any given phase. We can write from (\ref{chempot}) that $\beta\mu = \ln(v\rho)$. This can in turn be written $\mu = \partial f/\partial \rho$, where
\begin{equation}
\beta f = \rho(\ln\rho-1) +\int^\rho \ln v(s)\, \d s
\end{equation}
defines an effective free energy density $f(\rho)$ for homogeneous systems. This has a spinodal region if $v$ is a sufficiently decreasing function of $\rho$, specifically when 
\begin{equation}
\beta f''  = \frac{1}{\rho} + \frac{1}{v}\frac{dv}{d\rho} < 0.
\end{equation}
The spinodal, as usual, identifies a regime where the system is locally unstable to phase separation. One expect this to lie within a region of global instability whose boundaries are the binodal curve. 
For an equilibrium system, the binodal curve is found from $f(\rho)$ via the common tangent construction or CTC (Section \ref{CTCsection} above). It is tempting, but incorrect, to assume the same construction works in MIPS. To see why this fails, we need to consider the gradient corrections to the mean-field approach. In equilibrium, the gradient corrections to $\mu$ are constrained by the fact that $\mu = \delta F/\delta\rho$, with $F$ a free energy functional. Even though the CTC itself makes no mention of gradient corrections, it fails generally when these are neither zero nor of the equilibrium form. This leads to {\em anomalous phase coexistence} as described in Section \ref{APC} below.

We next therefore develop a square-gradient-level theory for MIPS, by supposing that $v(\r,[\rho]) \simeq v(\bar\rho(\r))$ where $\bar\rho(\r) = \int \rho(\r')K(\r-\r')\, \d\r'$ with $K$ a smooth kernel set by the range and form of the quorum sensing (or other) interactions present. Expanding the integral in gradients we find
\begin{eqnarray}
\bar\rho &=& \rho(\r) + {\bf c}_1\bdot \nabla\rho(\r) + c_2\nabla^2\rho(\r)\,,\\
{\bf c}_1 &=& \int {\bf s} K({\bf s}) \d{\bf s} = 0\,,\label{sgmips2}\\
c_2 &=& \frac{1}{2}\int s^2K({\bf s}) \d{\bf s}\,.
\end{eqnarray}
In (\ref{sgmips2}) we assumed an isotropic kernel: $K({\bf s}) = K(s)$.
Now Taylor exanding, we find $v(\bar\rho) = v(\rho) + c_2(v) dv/d\rho\, \nabla^2 \rho$, which gives
\begin{equation}
\mu = \ln\rho + \ln v + \frac{1}{v}\frac{dv}{d\rho} c_2(v)\, \nabla^2\rho\,.
\end{equation}
Note that $c_2$ can depend on $v$ (and hence on $\rho$) because the range of the quorum sensing can depend on how fast particles explore their surroundings by self-propulsion. 
Thus we have an effective chemical potential of the form $\mu = \mu_{\rm loc} - \kappa(\rho)\nabla^2\rho$.  We can compare this with the form arising from an equilibrium model such as $F = \int \left[f(\rho) + \kappa(\rho)(\nabla\rho)^2/2\right]\,\d\r$,
for which $\mu = \delta F/\delta\rho$ obeys
\begin{equation}
\mu = \frac{\delta F}{\delta\rho} = \frac{\partial f}{\partial \rho} + \frac{1}{2}\kappa'(\rho)(\nabla\rho)^2 - \del\bdot(\kappa(\rho)\del\rho) = \mu_{\rm loc} - \kappa(\rho)\nabla^2\rho - \frac{1}{2}\kappa'(\rho)(\nabla\rho)^2\,.
\end{equation}
Put differently, our active field theory has
\begin{equation}
\mu = \mu_{\rm passive}(\rho,\nabla\rho) + \frac{1}{2}\kappa'(\rho)(\nabla\rho)^2 
\end{equation}
whose last term is an explicit violation of TRS at square-gradient level. 

At the same level one could introduce corrections to the diffusivity $D = D_{\rm loc} + D_1(\rho) \nabla^2\rho$. However, this correction enters the diffusivity and the noise together, so that the FDT is maintained. We therefore ignore this quasi-passive correction, and conclude that the leading-order effect of activity in the quorum-sensing model enters purely via the fact that $\mu \neq \delta F/\delta\rho$. 
As we will see in Section \ref{ch:six}, more general models show other contributions at the same order. This exposes one of the disadvantages of the bottom-up coarse-graining approach, which is that the chosen microscopic model may not capture all the important physics of the general case.

\subsection{Anomalous phase coexistence}\label{APC}
We are now in a position to find the binodals for our MIPS theory and hence elucidate its anomalous phase coexistence. We proceed at mean-field level, without noise, allowing for a slightly more general form than found above:
\begin{eqnarray}
\dot\rho &=& -\del\bdot\J = \del\bdot M\nabla\mu\,, \\
\mu &=& \mu_0(\rho) + \lambda(\rho)(\nabla\rho)^2 - \kappa(\rho)\nabla^2\rho\,.
\end{eqnarray}
If an effective free energy $F$ exists, $2\lambda + \kappa' = 0$. We need to solve these equations in steady state ($\dot \rho = 0$) to find a density profile $\rho(x)$ along a one-dimensional coordinate $x$, normal to the interface between phases, that connects liquid and gas densities $\rho_{l,g}$. These coexisting densities can be viewed as eigenvalues of the nonlinear ordinary differential equation $\partial_x(M\partial_x\mu) = 0$; this has no solution if $\rho_{l,g}$ are not chosen correctly.

To solve this problem we follow~\cite{Solonphase}. First we set $\nabla\mu = 0$ which ensures no current (as is necessary to maintain static coexistence in a system with fixed boundaries). This requires $\mu_0(\rho_l) = \mu_0(\rho_g) = \bar\mu$, a constant. So one half of the CTC (equality of bulk chemical potentials) is sustained here. The other half (equality of the thermodynamically defined pressure, $P = \rho\mu-f$) is not, as we now show. We introduce a `pseudodensity' $R(\rho)$ obeying (with prime denoting $\rho$-derivative)
\begin{equation}
\kappa R'' = -(2\lambda+\kappa')R'\,.\label{trans1}
\end{equation}
Since $2\lambda +\kappa' = 0$ in passive systems, $R = \rho$ there (up to a constant) and we recover the standard construction in that case only. We also introduce a `pseudopotential' or transformed free energy $\Phi(R)$ obeying 
\begin{equation}
\frac{d\Phi}{dR} = \mu_0(\rho)\,.\label{trans2}
\end{equation}
We now notice that 
\begin{equation}
\int_{x_g}^{x_l} \mu \partial_x R\, \d x = \int_{x_g}^{x_l} \bar\mu \partial_x R\, \d x  = (R_l-R_g)\bar\mu\,.
\end{equation}
The left hand side can now be written using (\ref{trans2}) as
\begin{equation}
\int_{x_g}^{x_l} \mu_0 \partial_x R\, \d x -\int_{x_g}^{x_l} \partial_x\left[\kappa R'(\partial_x\rho)^2/2\right] \, \d x  = \Phi(R_l)-\Phi(R_g)\,.
\end{equation}
The choice of transformed variable now becomes clear: we have converted the gradient terms in this expression into a total derivative, causing them to disappear on integration. In the equilibrium case they are already a total derivative so the transform is not needed. Using (\ref{trans1}) we arrive at (with prime now denoting $R$ derivative)
\begin{equation}
\tilde P_g \equiv (\Phi'R-\Phi)_g = (\Phi'R-\Phi)_l \equiv\tilde P_l\,.
\end{equation}

Thus we have established equality of the `pseudo-pressure' $\tilde P$ between the phases, alongside equality of the chemical potential $\mu_0 = \Phi'$ as derived previously. This means that the binodals are, in general, found by applying the common tangent construction (CTC) to the function $\Phi(R)$ rather than $f(\rho)$ -- with equivalence in the case where the gradient terms derive from a free energy functional.
Translating back into the $\rho$ variable, one still has equality of the slope ($\mu = f'(\rho)$) in the two phases, but not of the thermodynamic pressure $\rho\mu-f$. This combination is sometimes called the `uncommon tangent construction'~\cite{Wittkowski14}. Note, in addition, that the thermodynamic pressure defined this way is only equal to the mechanical pressure (the force density on a bounding wall) for equilibrium systems. There are theories of the mechanical pressure for active particles but they lie beyond the scope of these lectures ({\em e.g.},~\cite{Solon15}).

\subsection{Minimal MIPS: Active Model B}
\begin{figure}
	\centering
	\includegraphics[width=5cm]{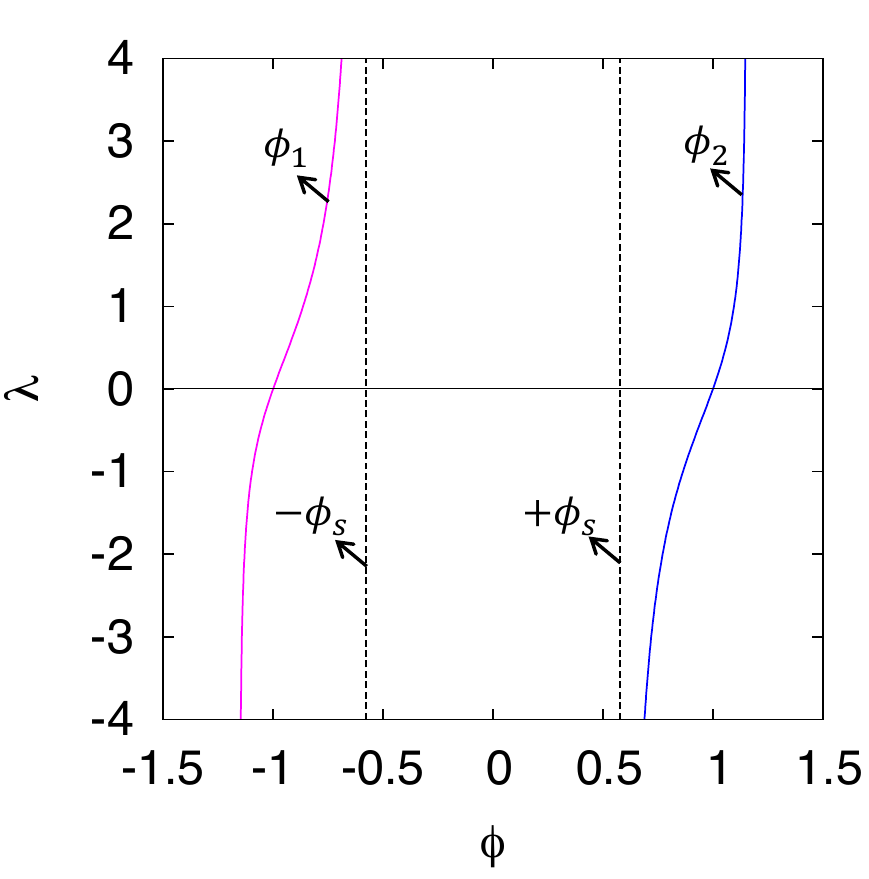}
	\caption[]{Binodals, which are $\lambda$-dependent, and spinodals (dotted), which are not,  for Active Model B (with $-a=b=\kappa = 1$). The system is globally unstable between the binodals and, in addition, locally unstable between the spinodals.}
	\label{AMBphasediag}
\end{figure}

Above we derived from microscopics a field theory involving a local effective free energy density $f(\rho)$ with a two-well structure (for sharply enough decreasing $v(\rho)$) but with gradient terms that destroy the free energy structure and break TRS. We connect this now with a top-down approach in which a minimal TRS violation is added to Model B, introduced in Section \ref{modB}. Guided by the quorum sensing model just discussed, we introduce Active Model B as:
\begin{eqnarray}
\dot\phi &=& -\del.\J\label{AModelB}\,,\\
\J &=& -\del\mu +\sqrt{2D}\bLambda\,, \label{AModelBJ}\\
\mu&=& \frac{\delta F}{\delta\phi} +\mu_A= a\phi + b\phi^3 -\kappa\delsq \phi +\mu_A\,,\label{AModelBmu}
\end{eqnarray}
where we have set the mobility $M$ to unity and introduced an active contribution to the chemical potential 
\begin{equation}
\mu_A = \lambda(\nabla\phi)^2\,.
\label{activemu}
\end{equation}
In the notation of the previous section, we have $2\lambda +\kappa' = 2\lambda \neq 0$ so that this model gives anomalous phase coexistence. The binodals were first calculated in~\cite{Wittkowski14}, by a somewhat less general method to the one given above. One finds that the two tangents to the symmetric $f(\phi)$ curve have intercepts differing by a term $\Delta P(\lambda)$ which is linear in $\lambda$ for small values but saturates at large ones. At large activity, the binodal closely approaches the spinodal on one side of the phase diagram (the negative $\phi$ side if $\lambda > 0$) but remains far from it on the other side. The resulting phase diagram (for $-a = b  = \kappa = 1$) is shown in Fig.~\ref{AMBphasediag}. 

Apart from anomalous phase coexistence, Active Model B (AMB) shows behaviour that quite closely resembles its passive counterpart~\cite{Wittkowski14}.  Specifically, it shows phase-separation kinetics with a coarsening exponent such that the typical length scale $L\sim t^\alpha$ with $\alpha$ close to the passive value of $1/3$. There is no sign of saturation at a finite value of $L$ as would be needed to explain microphase separation of the kind reported in cluster phases of self-propelling colloids~\cite{Buttinoni13}  and for simulations in ABPs with collisions, which show finite vapour bubbles surrounded by a dense liquid~\cite{Stenhammar13}.

An additional feature, or arguably a bug, of AMB is that, there is no possibility of nonzero-mean circulating real-space particle currents in steady state, although such currents are seen in many active systems such as bacteria in confined microfluidic spaces, and in simulations of ABPs with collisions in ratchet-like environments~\cite{Stenhammar16}. The reason for this is simple: the deterministic current in the model is of pure gradient form, so that $\langle \J\rangle = -\nabla\mu$ and $\nabla\times\langle\J\rangle = \0$. 

\section{Entropy production in active field theories}\label{EPAFT}\label{ch:four}
In view of the above discussion, one might even wonder whether AMB is active at all: could there be some dynamical generalization of the `pseudo-transform' that maps the entire dynamics onto some passive system? The answer is no, and in this Section we give a formal procedure for deciding such questions. Specifically, we introduce an objective quantification of steady-state irreversibility using the tools of stochastic thermodynamics. For a review of that field, see~\cite{Seifert}. One of its main results is a theorem concerning the steady-state entropy production rate $\cal S$:
\begin{equation}
{\cal S} = \lim_{\tau\to \infty} \,\frac{1}{\tau}{\Big\langle}\ln \frac{{\mathbb P}_F[\Psi(\r,t)]}{{\mathbb P}_B[\Psi(\r,t)]}{\Big\rangle}\,.
\end{equation}
Here the subscripts on the path weight refer to forward and backward paths as previously considered in Section \ref{actpass}; angle brackets are an ensemble (or noise) average. 

To set the stage, we return to the passive example of Model A. For that model we found in (\ref{ratioweight},\ref{ratioweight2})
\begin{equation}
\ln \frac{{\mathbb P}_F[\Psi(\r,t)]}{{\mathbb P}_B[\Psi(\r,t)]} = \beta\int_1^2\dot\phi\frac{\delta F}{\delta\phi} \, \d\r\, \d t = \beta (F_2-F_1)\,, \label{SMA}
\end{equation}
so that ${\cal S} = \lim_{\tau\to \infty} \beta(F_2=F_1)/\tau = 0$ because $F$ cannot change without bound in steady state.

For AMB we have $\dot\phi = -\del\bdot(-\nabla\mu + \sqrt{2D}\bLambda)$ or $\dot\phi -\nabla^2\mu = -\sqrt{2D}\,\del\bdot\bLambda$. Recalling (\ref{whiteweight}) we construct
\begin{equation}
{\mathbb P}_{F,B} = {\cal N}_\phi \exp\left[-\frac{1}{4D}\int[\nabla^{-1}(\pm\dot\phi-\nabla^2\mu)]^2\, \d\r \d t\right]  = {\cal N}_\phi \exp[-{\cal A}_{F,B}]\,, \label{ambweights}
\end{equation}
where $\nabla^{-1}$ is the inverse of the divergence operator (we define this below) and the only difference between forward ($+$) and backward ($-$) weights  is the sign of $\dot\phi$. The forward and backward actions obey $4D{\cal A}_{F,B} = \int{\mathbb A}_{F,B} \, \d\r \,\d t$ where,  after integration by parts, we find for the action densities
\begin{equation}
{\mathbb A}_{F,B} = - (\pm\dot\phi-\nabla^2\mu)\nabla^{-2}(\pm\dot\phi-\nabla^2\mu)\,.
\end{equation}
Here $\nabla^{-2}$ is the inverse of the Laplacian, {\em i.e.,} an integral against its Green's function, so that in 3D one has $\nabla^{-2}\delta(\r') = {-1}/{4\pi|\r-\r'|}$. This also allows us to make sense of the formal operator $\nabla^{-1}$ in (\ref{ambweights}) by defining $\nabla^{-1}X = \nabla\nabla^{-2}X$. 

Next we construct
\begin{equation}
\cal S = \lim_{\tau\to \infty}\frac{1}{4D\tau}\int\langle{\mathbb A}_B-{\mathbb A}_F\rangle \d \r\, \d t  = -\lim_{\tau\to \infty}\frac{1}{D\tau}\int\langle\mu_A(\r,t)\dot\phi(\r,t)\rangle \d \r\, \d t \,,
\end{equation}
where a boundary term involving $\int\dot\phi\delta F/\delta\phi$ vanishes for the same reason as in equilibrium (see (\ref{SMA})). The angle brackets denote ensemble averages over the stationary measure of the theory; in situations such as phase separation which break ergodicity, this average is taken at fixed interfacial positions. We can therefore write
\begin{equation}
{\cal S}  = \int \sigma_\phi(\r)\, \d\r\,,
\end{equation}
where
\begin{equation}
\sigma_\phi = -\frac{1}{D} \langle \mu_A\dot\phi\rangle(\r) = -\frac{1}{D} \langle \mu_A\nabla^2\mu\rangle(\r) \label{entprod}
\end{equation} 
can be viewed as a local entropy production rate. When symmetry is broken by phase separation, this depends on position, as the notation implies. 
In (\ref{entprod}) the second equality is found by substituting the equation of motion for $\dot\phi$ and noting that in the It\^{o} prescription, the contributions involving a noise average vanish.  The above arguments are discussed in~\cite{Nardini17} where they are derived in more detail. 

In calculating the entropy production in MIPS at mean-field level we first note that in the steady-state without noise, $\nabla^2\mu$ zero but so is $D$. Thus to evaluate $\sigma_\phi$ we need to take the low-noise limit in (\ref{entprod}) explicitly. One finds that $\phi = \phi_0+\sqrt{D}\phi_1$ where $\phi_0$ is either a uniform or a phase-separated density profile. The result~\cite{Nardini17} is that ${\cal S} = \int \sigma_\phi\,\d\r = c_1 I + c_2D V$ with $c_{1,2}$ numerical factors that depend on $\lambda$. Here  ${\cal S}$ contains a term independent of the noise level $D$ that is proportional to $I$, the area of the interface(s) between bulk phases. There is a second term, proportional to the bulk volume (with a different proportionality constant in each phase), that scales as $D$. Therefore in homogeneous phases, the steady state entropy production is finite but caused solely by fluctuations, which break TRS. In a phase separated state, there is a localized interfacial contribution to the entropy production even in the zero-noise limit.  This contribution can perhaps be viewed as arising from dissipative `pumping' across interfaces. Indeed, to explain anomalous phase separation in which bulk densities differ from those that would minimize the effective free energy terms, there has to be some mechanism that pumps material from one side of an interface to the other, with diffusive transport presumably balancing that in steady state. 

In any case, we have shown, as promised, that AMB is not equivalent to any passive model, because its steady-state entropy production rate ${\cal S}$ is finite. Any failure to explain generic active phenomena, such as microphase separation, is not caused by some hidden reversibility, but by over-simplifications of the model. For instance, as noted previously, there are no circulating real-space currents in steady state at deterministic level. This is rectified in an extended model called AMB+ which we discuss in Section \ref{ch:six} below. First we address a different question, which is how to couple AMB to a momentum-conserving solvent.

\section{Active Model H}\label{ch:five}
For systems of active particles (without orienting interactions) in which the fluid velocity field $\v$ plays an important role, the natural starting point at continuum level is Model H (see Section \ref{passiveH}). To this we can now add minimal TRS-breaking terms. One such term, in the chemical potential, is the $\lambda(\nabla\phi)^2$ term of AMB.
Another new term enters the NSE (\ref{HNSE}) whose passive version contains a  stress obeying $\del{\bdot}\bsigma = -\phi\del\mu$. This form assumes a thermodynamic relation between stress and chemical potential which holds for genuine equilibrium systems, but not for active ones even if they happen to have a free energy structure in the diffusive sector of the model (such as MIPS with $\lambda = 0$). Put differently, only in true equilibrium models do mechanical forces and thermodynamics stem from a shared microscopic Hamiltonian. In a system undergoing motility-induced phase separation, for instance, even the fact that the local ``free energy density" $f(\phi)$ has two minima can arise purely from activity and not from attractive interactions. This means that, while the quasi-passive terms in the equation of motion for the density $\phi$ do not break time-reversal symmetry in themselves, they have no reason to feed through via thermodynamics into the stress term in the NSE.

What matters in the NSE for an incompressible fluid is the deviatoric stress, which is traceless and differs from the full stress by a diagonal (pressure) contribution. From (\ref{stress3}), the deviatoric stress is 
\begin{equation}
\Sigma_{ij}^D = -\tilde\zeta\left((\partial_i\phi)(\partial_j\phi) - \frac{1}{d}|\del\phi|^2\delta_{ij}\right), \label{devio}
\end{equation}
in which $\tilde\zeta = \kappa$. At large enough scales this term encodes the presence of an effectively structureless interface between phases with a certain surface tension, whose dependence on the parameters of the model can readily be found \cite{CatesTjhungJFM}.

Without an external field to break rotational invariance, the form (\ref{devio}) is in fact the only possibility to leading order in gradients. Therefore, in generalizing from the passive to the active case, what we lose is the connection between $\tilde\zeta$ and $\kappa$.  Active Model H thus
reads \cite{Tiribocchi15}
\begin{eqnarray}
\rho(\dot\v+\v\bdot\del\v) &=& \eta\delsq \v-\del P - \del\bdot\bSigma^D +\del\bdot\bSigma^N,\label{AHNSE}\\
\del\bdot\v &=& 0, \label{Aincomp}\\
\dot\phi + \v\bdot\del\phi &=& -\del\bdot(-M\del\mu+\J^N),\label{AHphi}\\
\mu(\r) &=& a\phi+ b\phi^3 -\kappa\delsq \phi+\lambda|\del\phi|^2\,,\label{AmumodelH}
\end{eqnarray}
with $\bSigma^D$ obeying (\ref{devio}), in which $\tilde\zeta$ is now a parameter that depends on both interaction forces and activity. Because this is no longer linked to $\kappa$  in (\ref{AmumodelH}), which is always postive, $\tilde\zeta$ can have either sign. 

\begin{figure}
\centering
\includegraphics[height=4cm]{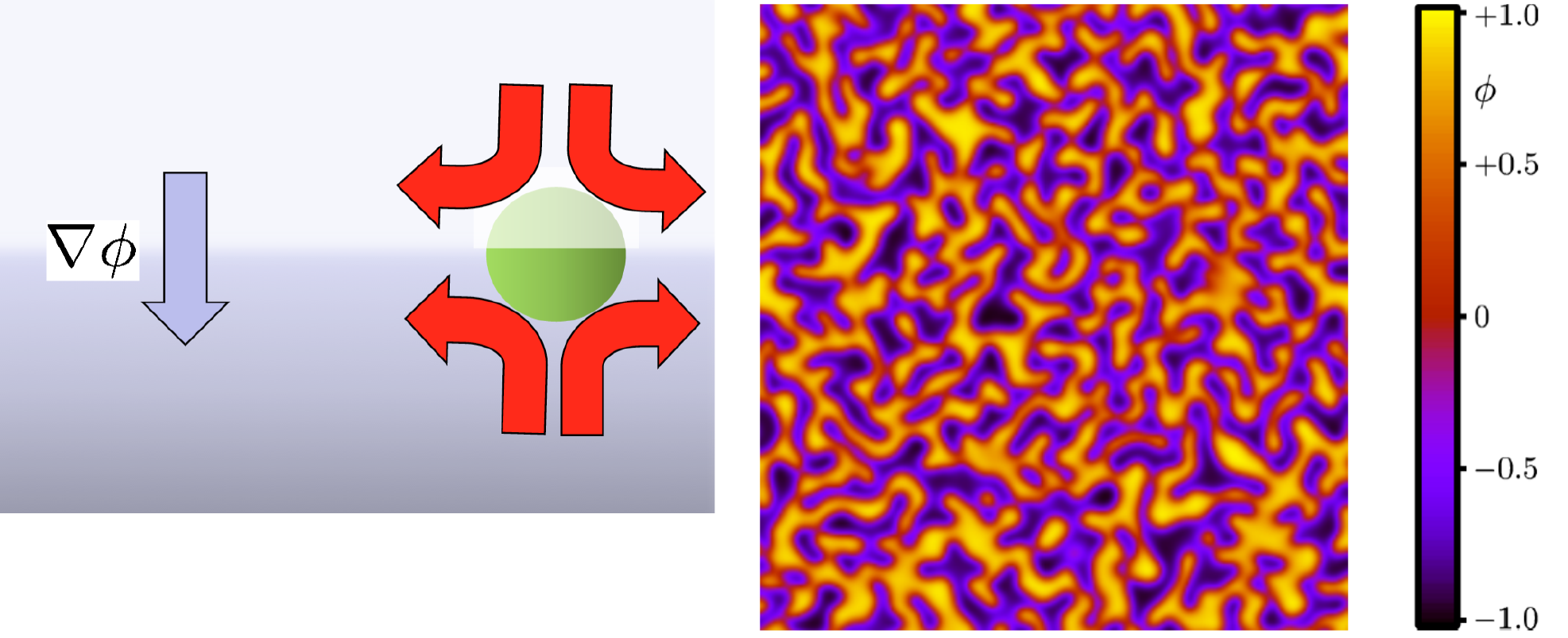}
\caption{Left: Schematic of stretching flow at an interface between low- and high-density regions ($\phi<0$ and $\phi>0$ respectively) of contractile swimmers. Right: Snapshot of a steady state in which this interfacial stretching balances the intrinsic coarsening dynamics of Model H. (Image courtesy of A. Tiribocchi.)}
\label{fig:AMH}
\end{figure}

Specifically, it includes an active contribution that is positive for extensile swimmers, and negative for contractile ones. Indeed, this is the same physics as described by (\ref{swimstressp}). This is not surprising since the adiabatic approximation for rotational relaxation of scalar particles implies a polarization $\p\sim\nabla\phi$ (see the discussion after (\ref{moment2}); substituting this into (\ref{swimstressp}) gives $-\zeta(\nabla\phi)(\nabla\phi)$ as the leading order active stress, and hence $\tilde\zeta = \zeta+\kappa$ in (\ref{devio}) -- see Fig.~\ref{fig:AMH}. Note, incidentally, that if one had some specific mechanism to align particles' propulsive directions in the plane of the interface, rather than perpendicular to it as implied by $\p\sim\nabla\phi$, the active contribution could get reversed in sign. For concreteness we ignore this case here.

Because $\zeta\neq\kappa$, there are effectively two interfacial tensions in Active Model H, one controlling the diffusive flux (set by $\kappa$) and one controlling the fluid flow, now set by $\tilde\zeta$. (Note that both effective tensions enter solely via mechanical stresses in the momentum balance equation (\ref{AHNSE}); these should not be confused with the pseudotension defined in Section \ref{ch:six} below.) Although this mismatch breaks TRS, if both tensions are positive the consequences appear relatively mild, at least according to numerical studies of the coarsening behaviour~\cite{Tiribocchi15}. On the other hand, for negative $\tilde\zeta$, as would arise for sufficiently contractile swimmers, there is effectively a negative interfacial tension in the mechanical sector. Swimming particles tend to orient perpendicular to the interface between phases, where their contractile action pulls fluid towards the interface and pushes it out in sideways causing the interface to stretch (Fig.~\ref{fig:AMH}, left), giving in effect a negative mechanical tension. In consequence, phase separation can arrest at a finite length scale where the diffusive shrinkage of the interfacial area is balanced by its contractile stretching (Fig.~\ref{fig:AMH}, right). 

This offers a hydrodynamic, rather than microscopic, mechanism for the existence of cluster phases, but only in cases where the swimming is contractile \cite{Tiribocchi15}. Moreover, experimental reports of arrested phase separation in active colloids generally concern systems where particles reside near the bottom wall of the container; this wall absorbs momentum resulting in a ``semi-dry'' situation to which Active Model H does not directly apply. A more microscopic approach involving simulation of colloidal swimmers with full hydrodynamic and chemical interactions beside a wall gives a somewhat different explanation of cluster phase formation~\cite{Singh18}. Indeed, most explanations of microphase separation in active scalar systems invoke some specific chemical or other interactions, which are indeed generally present as a by-product of the self-propulsion mechanism~\cite{Buttinoni13,Saha14,Liebchen}

\section{Active Model B+}\label{ch:six}
Active Model B, as presented above, with an active contribution to the chemical potential $\mu_A = \lambda(\nabla\phi)^2$, was directly inspired by a coarse-grained microscopic model for self-propelled particles with density-dependent speed. We now extend it to include a term that was not suggested by that particular microscopic model, but is of the same order in the expansion in $\nabla,\phi$ and therefore ought to be treated on the same footing. A third term of this order can be viewed as a density-dependent square gradient coefficient, $\kappa \to \kappa(0) + \kappa'\phi$. However, as shown in Section \ref{MFMIPS}, $\kappa'$ can effectively be absorbed into the activity coefficient $\lambda$, in the sense that any model with $2\lambda+\kappa'  = 0$ supports an effective free energy. As originally presented in~\cite{Nardini17} AMB+ also included further, higher order terms that we also ignore. 

For our purposes therefore, we can define AMB+ as
\begin{eqnarray}
\dot\phi &=& -\del.\J\label{AModelB+}\,,\\
\J &=& \J_D +\sqrt{2D}\bLambda\,, \label{AModelB+J}\\
\J_D &=& -\nabla\frac{\delta F}{\delta \phi} - \lambda\nabla((\nabla\phi)^2)
+\zeta(\nabla^2\phi)\nabla\phi \label{AModelB+JD}\,.
\end{eqnarray}
This differs from the previous AMB (\ref{AModelB}-\ref{AModelBmu}) solely by the term in $\zeta$.
Note that $\zeta$ here has no relation to the mechanical activity parameter introduced for wet polar liquid crystals and for Active Model H in previous sections; from now on we are discussing dry systems only. In other words, $\phi$ is the only order parameter, there is no Navier-Stokes sector, and (\ref{AModelB+}-\ref{AModelB+JD}) are complete as they stand. 

The new structure of (\ref{AModelB+JD}) ensures that the deterministic current $\J_D$ is no longer curl-free in general: circulating currents are possible. However in one dimension (only), we have $-\lambda\nabla((\nabla\phi)^2)
+\zeta(\nabla^2\phi)\nabla\phi = -\lambda_{\rm eff}\nabla((\nabla\phi)^2)$ where $\lambda_{\rm eff} = \lambda -\zeta/2$. In any strictly 1D geometry, therefore, the results of AMB still apply, subject to this shift in $\lambda$.
Intriguingly, such geometries encompass the mean-field calculations of bulk phase equilibria, including the anomalous coexistence discussed in Section \ref{APC}. This is because such calculations assume an interfacial profile $\phi(x)$ that depends on a single normal coordinate. Therefore at strictly mean field level the phase diagram of AMB+ is identical to the one shown in Fig.~\ref{AMBphasediag} with $\lambda$ replaced by $\lambda_{\rm eff}$ on the vertical axis.

\begin{figure}
\centering
\includegraphics[height=6cm]{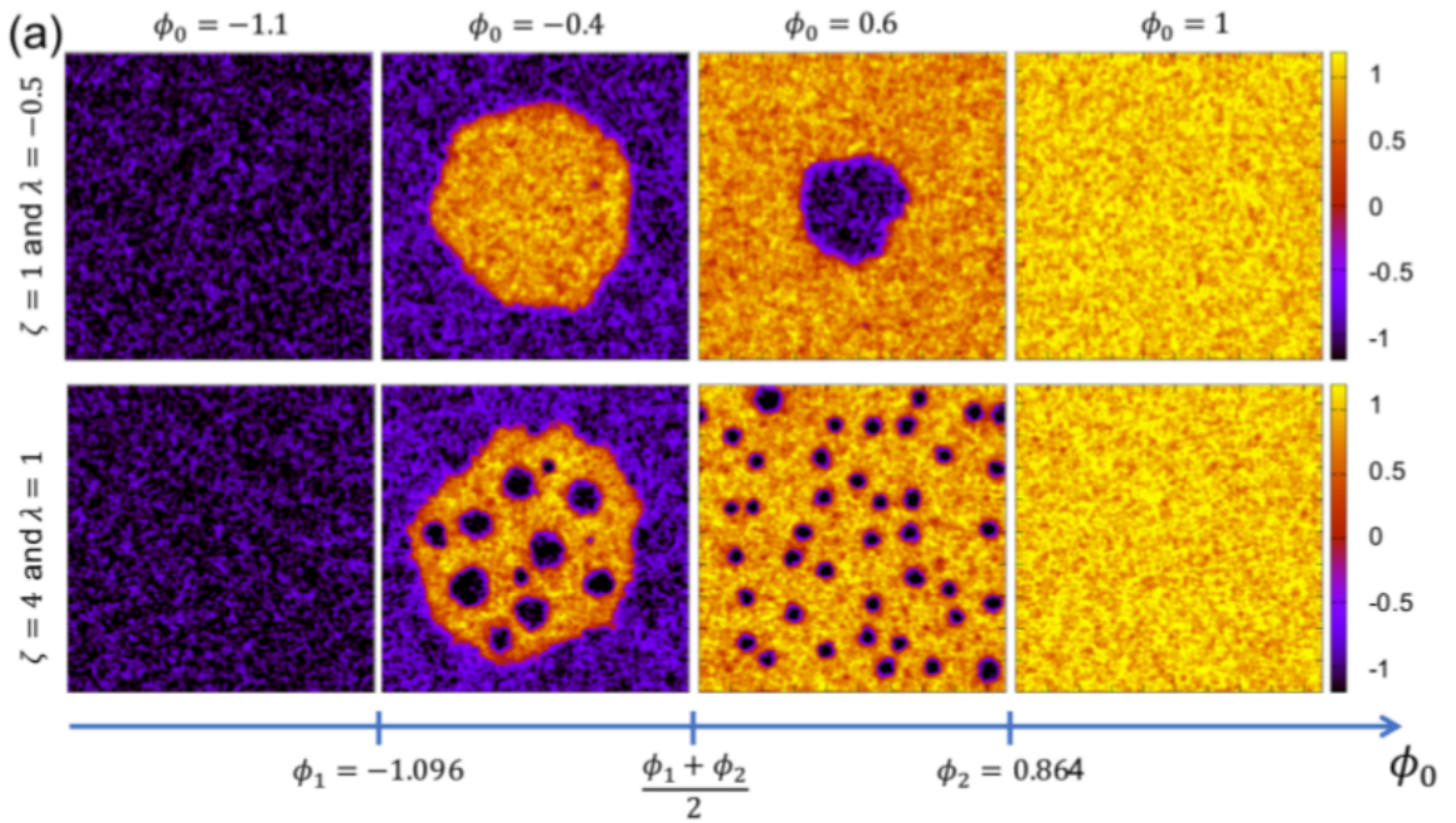}
\caption{Sequence of phase equilibria in AMB+on varying the global density $\phi_0$ for two $\zeta$ values at the same $\lambda_{\rm eff}$; here $\phi_{1,2}$ are the binodals. In this finite-sized system bulk phase coexistence appears as a large droplet of one phase in the other. The larger $\zeta$ value (bottom row) shows coexistence between a microphase separated bubble phase and excess vapour, and also a space-filling bubble phase (Image courtesy of E. Tjhung.) }
\label{AMB+phasediag}
\end{figure}

On the other hand, if one simulates numerically AMB+ at finite noise levels one discovers two facts~\cite{Tjhung18}. First, the mean-field predictions are robust to (moderate) noise so long as $\zeta = 0$ (the pure AMB case), but second, there is a region of microphase separation in cases where $\zeta$ is large enough in magnitude. For positive $\zeta$ this is the narrow region between the spinodal and the binodal in the bottom right of the phase diagram of Fig.~\ref{AMBphasediag}. Here one sees droplets of vapour surrounded by dense liquid. Moreover for $\phi$ lying between this spinodal and the opposite binodal -- the range in which bulk phase separation might be expected -- one has coexistence of this state of bubbles with an excess phase of bulk vapour.  This kind of `bubbly phase separation' appears to arise in active Brownian particles with hard-core collisions~\cite{Stenhammar13} and is shown for AMB+ in Fig.~\ref{AMB+phasediag}. Note also the AMB+ is symmetric under $\lambda,\zeta,\phi\to -\lambda,-\zeta,-\phi$. Hence for negative $\zeta$, the identity of dense and dilute phases gets reversed, so the microphase separated region lies between the binodal and spinodal at the the top left in Fig.~\ref{AMBphasediag}.  Here dense liquid droplets are surrounded by vapour. We interpret this as a cluster phase, which again coexists with bulk excess liquid when the mean density is pushed further into the miscibility gap.

\begin{figure}
\centering
\includegraphics[height=4cm]{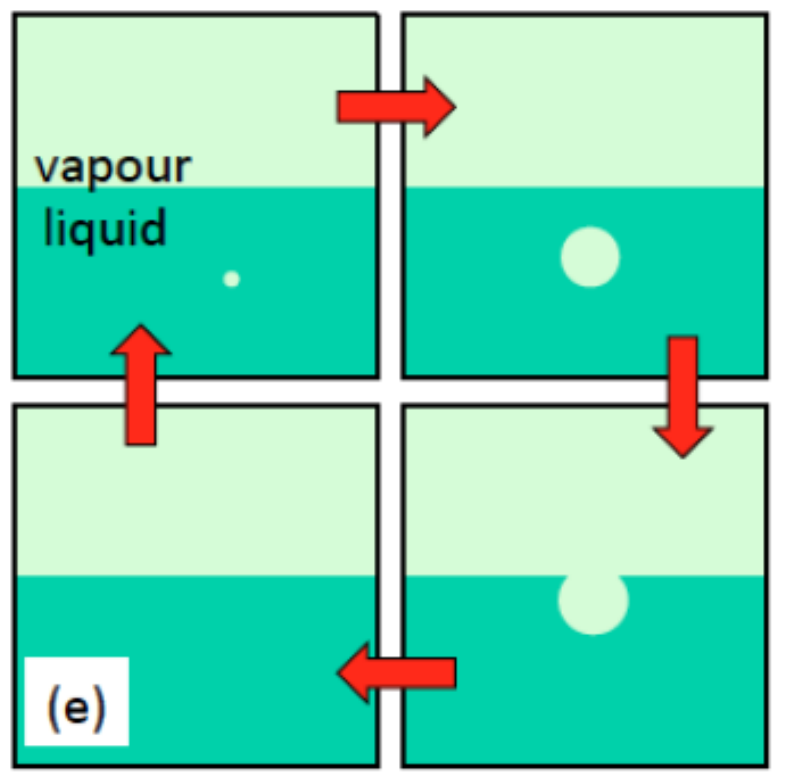}
\caption{Life-cycle of bubbles near the interface with a bulk vapour phase. Bubbles nucleate, grow, drift to the interface, and pop. The reverse process (engulfment of a pocket of vapour, followed by its drift away from the interface, shrinkage and disappearance) is not seen.}
\label{lifecycle}
\end{figure}

As stated already, the $\zeta$ term allows $\del\times\J_D \neq \0$ by breaking the pure gradient structure of the deterministic current. We have just seen that it also promotes microphase separation where bulk phase separation would otherwise be seen;  we explain this below.  The $\zeta$ term also has a third effect, which is to create pronounced circulating currents in phase space. For a state where a bubble phase coexists with excess vapour, these take the form of the life-cycle for bubbles shown in Fig.~\ref{lifecycle}.

\subsection{Coarsening dynamics}
To understand the origin of the microphase-separated states in AMB+, we must first recall the kinetic pathway that leads to bulk phase separation in passive Model B. 
We start by considering a droplet of dense liquid in equilibrium with a finite environment of vapour. (We could have chosen a vapour droplet surrounded by liquid, but this is the more conventional language.) This state is indefinitely stable in a finite system so long as the droplet has a lower interfacial area than a slab of liquid, of equal volume, with flat interfaces.   Recall that for flat interfaces in a passive system, $\mu_1 = \mu_2$ and $P_1=P_2$ where $\mu = \partial f/\partial\phi$ and $P = \mu\phi-f$ are the bulk chemical potentials of the phases involved. For a droplet geometry, the resulting binodals $\phi_{1,2} = \pm\phi_B$ are subject to corrections of order $1/R$ that we now calculate.

The pressure inside our droplet is greater than that outside by an amount $\Delta P$ because there is a finite surface tension at the interface between phases. (We denote this by $\sigma$ and will not need its value, although this is well known for Model B \cite{CatesTjhungJFM}.) To calculate $\Delta P$ note that the total force on the upper half of the droplet exerted by the bottom half is (in three dimensions)
$\pi R^2\Delta P -2\pi \sigma R = 0$, which must vanish if the droplet is not moving. The first term comes from the resolved vertical component of the outward pressure jump acting across the droplet surface, and the second is the surface tension force acting across its perimeter. The resulting pressure excess is known as `Laplace pressure'. In $d$ dimensions the generalization is
\begin{equation}
P_2 = P_1 +\frac{(d-1)\sigma}{R}\,.\label{LP}
\end{equation}

In static equilibrium we must have $\J = \0$ so that $\nabla\mu=0$, requiring $\mu_1=\mu_2$ just as for flat interfaces. Allowing for $\Delta P$ we thus have an uncommon tangent construction on $f(\phi)$ in Fig.~\ref{one}, with the intercepts of our two tangents differing by $\Delta P$. (To get the sign right, both tangents must be drawn sloping upwards to the right on that figure.) For large $R$ we can expand $f(\phi)$ to quadratic order around each of its minima at $\pm\phi_B$. Since the curvature at each minimum is the same ($f''(\phi_B) = f''(-\phi_B) \equiv \alpha$) we have $\phi_1 = -\phi_B+\Delta$, $\phi_2 = \phi_B+\Delta$ and $\mu_1 =\mu_2 = \alpha\Delta$. Moreover, by an easy calculation, $P_1-P_2 = -2\alpha\phi_B\Delta$ so that
\begin{equation}
\Delta = \frac{d-1}{2f''(\phi_B)\phi_B}\,\frac{\sigma}{R} + O(R^{-2})\,.
\end{equation}
This formula tells us how the coexisting densities across the interface of a droplet differ from the equilibrium binodals; specifically, just outside a droplet of radius $R$,  $\phi$ is raised by a `supersaturation' $\Delta(R)$.  If one has multiple droplets, these supersaturations are not matched in general, and the result is a diffusive flux from smaller to larger droplets.

\begin{figure}
	\centering
	\includegraphics[width=7cm]{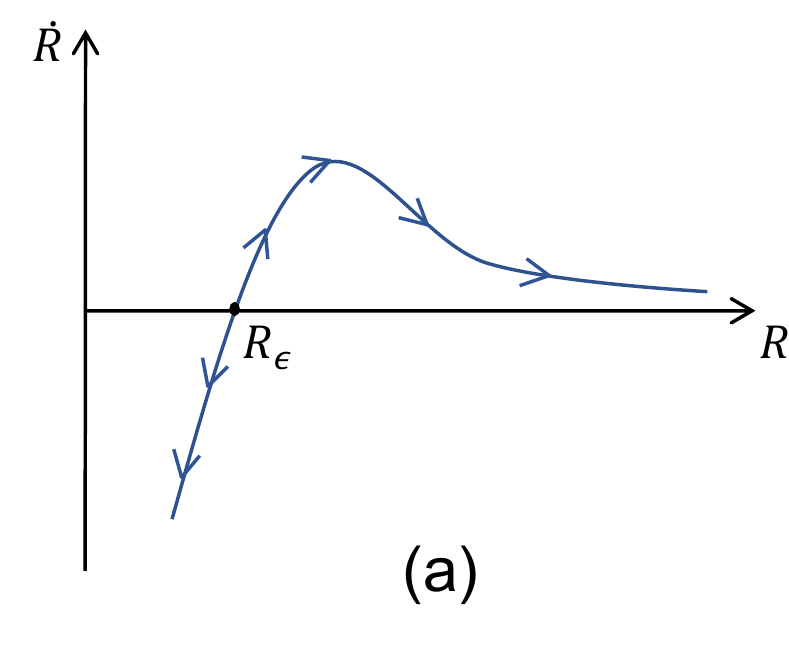}
	\caption[]{Growth rate of a droplet as a function of size during the Ostwald process.}
	\label{three}
\end{figure}

To see this, we take a mean-field approach to the multi-droplet situation in which a bath of other droplets is represented by a distant reservoir of supersaturation $\epsilon$. We then seek a spherically symmetric quasi-static ($\dot\phi = -\del\bdot\J = 0$) exterior solution, $\phi(r) = -\phi_B+\tilde\phi(r)$, where $\tilde\phi(R) = \Delta$ and $\tilde\phi(\infty) = \epsilon$. With $\J = - \del\mu$ this solution obeys $\mu = \alpha\tilde\phi$ where $\nabla^2\tilde\phi = 0$; note that the square gradient contribution to $\mu$ vanishes in this geometry. With the given boundary conditions, the result is $\tilde\phi = \epsilon +(\Delta-\epsilon)R/r$. If we now calculate the radial current $J = |\J|$ exterior to the droplet, we have $J = -\alpha\partial\tilde\phi/\partial r = \alpha(\Delta-\epsilon)R/r^2$; just outside the droplet itself this becomes $-\alpha(\Delta-\epsilon)R$. By mass conservation (noting that $\phi$ jumps by $2\phi_B$ across the interface) we then find $\dot R = -J(R)/2\phi_B$ or
\begin{equation}
\dot R = \frac{1}{2\phi_B}\left[\frac{\alpha}{R}\left(\epsilon - \Delta(R)\right)\right] = \frac{1}{2\phi_B}\left[\frac{\alpha}{R}\left(\epsilon - \frac{\sigma}{\alpha\phi_BR}\right)\right].
\label{evaporation}
\end{equation}
This behaviour is sketched in Fig.~\ref{three}.

The result of this dynamics is Ostwald ripening, in which material is transported from small droplets to large ones by diffusion. Assuming that the system has not already reached its end-point of full phase separation, the ambient supersaturation $\epsilon$ in remains positive, and $\dot R(R)$ exhibits an unstable fixed point at 
\begin{equation}
R = R_\epsilon(t)\equiv \frac{\sigma}{\alpha\phi_B\epsilon}.
\end{equation}
Droplets bigger than this grow, those smaller, shrink. 
We can find the scaling of the typical droplet size $\bar R$ by assuming this to be comparable (but not equal) to $R_\epsilon$ and conversely that the ambient supersaturation $\epsilon$ is comparable (but not equal) to the local supersaturation near a typical drop: $\epsilon \simeq \Delta(\bar R) = \sigma/\alpha \bar R\phi_B$. Substituting in (\ref{evaporation}) gives
\begin{equation}
\dot {\bar R} \simeq \frac{\alpha\sigma}{\phi_B\bar R^2},
\end{equation}
which results in the scaling law $R(t)\sim t^{1/3}$. 
For a more complete theory of Ostwald ripening, known as the Lifshitz-Slyov-Wagner theory, see~\cite{Onuki02}. 

\subsection{Reverse Ostwald process in AMB+}
The Ostwald process described above is broadly unchanged in Active Model B (without the +). Although that model shows an uncommon tangent construction even for flat interfaces, the effect of this, and the one arising from Laplace pressure, are essentially additive~\cite{Wittkowski14}. Hence all supersaturations are increased by a fixed, activity-dependent amount and coarsening proceeds normally: activity doesn't change the diffusive competition between large and small droplets driven by the curvature dependent Laplace pressure terms.

AMB+ is different from this, as we now show by writing the deterministic part of the current as 
\begin{eqnarray}
\J_D &=& -\nabla(\mu_E+\mu_A) + \J_\zeta\,,\\
\mu_E &=& a\phi+b\phi^3 -\kappa\nabla^2\phi\,,\label{amb+muE}\\
\mu_A &=& \lambda(\nabla\phi)^2\,, \\
\J_\zeta &=& - \nabla\mu_\zeta + \del\times\A\,.
\end{eqnarray}
Here the current arising from $\zeta$ has been formally subject to Helmholtz decomposition into its pure gradient and pure curl parts.
Its pure curl part, $\del\times\A$, while conceptually very important when thinking about circulating currents, has no effect whatever on $\dot\phi$. Indeed we have
\begin{equation}
\dot\phi = \nabla^2\mu = \nabla^2(\mu_E+\mu_A+\mu_\zeta)\,.\label{fullmu}
\end{equation}
Given this structure, one might wonder how the standard Ostwald calculation could ever be circumvented. Importantly however, the Helmholtz decomposition delivers a contribution to the chemical potential that is nonlocal:
\begin{equation}
\mu_\zeta(\r) = - \nabla^{-2} \del\bdot\J_\zeta\,,
\end{equation}
where the integral operator $\nabla^{-2}$ was defined (via the Greens function of the Laplacian) and discussed in Section \ref{EPAFT}.

Mathematically there is a direct analogy between calculating $\mu_\zeta$ and the electrostatics problem of calculating the potential $V$ caused by a charge density $-\del\bdot\J_\zeta$; $\J_\zeta$ is then analogue to the electric field. The problem is particularly simple for a spherically symmetric droplet since the field $E(r)\hat\r$ is radial and $V$ likewise is a function of $r$ only. Moreover $V(r) = -\int_r^\infty E(s)\,\d s$. Following directly this analogy, one finds~\cite{Tjhung18}
\begin{equation}
\mu_\zeta(r) = \zeta\int_r^\infty (\nabla^2\phi)\nabla\phi\, \d s  = 
-\frac{\zeta}{2}\phi'(r)^2 + (d-1)\zeta\int_r^\infty \frac{\phi'(s)^2}{s} \d s\,.
\end{equation}
The second term is the nonlocal part. For a droplet with a sharp interface (on which all the `charge' resides) this term is zero inside the droplet, jumping across the interface to an exterior value that is well approximated by
\begin{equation}
\Delta_\zeta \simeq (d-1)\frac{\zeta}{R}\int_{-\infty}^{\infty} \phi'(s)^2\, \d s\,.
\end{equation}
The full chemical potential $\mu$ in (\ref{fullmu}) must however be continuous across the interface, just as it is in the passive and AMB cases. Accordingly we must have
\begin{equation}
a\phi_1+b\phi_1^3 = a\phi_2+b\phi_2^3+ \Delta_\zeta
\end{equation}
so that the {\em bulk} values of $\mu_E$ either side of the interface (found by setting $\nabla\phi = 0$ in (\ref{amb+muE})) are now unequal. This means that, for the first time in these lecture notes, there is not even an uncommon tangent construction to find the supersaturation as a function of droplet size.

To proceed in this unusual situation, we return to the transform that allowed us to calculate anomalous phase coexistence for flat interfaces. In a notation more suited to the present context (where $R$ is droplet radius so cannot be used for the transformed density), we introduce
\begin{eqnarray}
\varphi(\phi)& : \;\;& \kappa\varphi'' = (\zeta-2\lambda)\varphi'\,,\\
\Phi(\varphi)& : \;\;& \frac{d\Phi}{d\varphi} = f'(\phi)\,,\\
\tilde P &= &\Phi'\varphi-\Phi\,.
\end{eqnarray}
By again writing $\mu$ in two forms (one explicitly constant and the other the combination of variables that must equal this constant), multiplying by $\partial_r\varphi$ and integrating, one arrives at the analogue of the Laplace pressure equation for the pseudopressure $\tilde P$:
\begin{equation}
\tilde P(\phi_1) = \tilde P(\phi_2) +\frac{d-1}{R}\tilde\sigma\,,
\end{equation}
where $\tilde\sigma$ is a pseudotension. Although the interface is no longer flat, these results can be established because $\phi$ is still a function of a single (radial) variable~\cite{Tjhung18}.

The remarkable final twist in the tale, whose details can be found in~\cite{Tjhung18}, is that {\em the pseudotension is negative} for large negative $\zeta$. Conversely, if one makes the same calculation for a vapour droplet surrounded by liquid, it is negative for large positive $\zeta$. These are precisely the regimes where microphase separation was seen numerically in AMB+ for cluster and bubble phases respectively (compare Fig. \ref{AMB+phasediag}).  So far we have not found a clear physical interpretation for the negative pseudotension. Indeed we don't have a clear interpretation for any of the pseudo-quantities; nonetheless, these allow us to make headway in calculating phase equilibria for active systems by converting their mathematics into a form made familiar by a century of work on equilibrium thermodynamics.

If we accept that the pseudotension in the Ostwald calculation is negative, microphase separation is a natural consequence. Whatever distribution of droplet sizes one starts with, large droplets shrink and small ones grow because the arrows on Fig.~\ref{three}, where the vertical axis is now $-\dot R$ rather than $\dot R$, get reversed, giving a {\em stable} fixed point at $R_\epsilon$. All droplets then converge on a fixed size that is set by the initial conditions. This is an interesting outcome for two reasons: first it gives a natural and generic explanation of microphase separation in scalar active field theories,  and second it says that the reverse Ostwald dynamics does not select any particular length scale for the microphase-separated state. Numerical work indicates that in practice this length scale is selected not by initial conditions but by the noise level, through a mechanism that is not yet fully characterized~\cite{Tjhung18}. 

\section{Conclusion and outlook}\label{ch:seven}
In these lectures I started by surveyed some of the traditional field theories of soft condensed matter, and then addressed some of the active field theories used to describe at continuum level systems of self-propelled particles of various types. Particular attention was paid to the case of particles which have no alignment interactions and can therefore be described by a conserved scalar field. A theory of this field, for systems without coupling to a momentum-conserving solvent, was explicitly constructed by coarse-graining a carefully chosen microscopic model. This theory describes MIPS (motility-induced phase separation).  The resulting theory was then simplified to a canonical, $\phi^4$ form (Active Model B) and its anomalous phase separation discussed. After a brief excursion into the momentum-conserving case (Active Model H), we returned to Active Model B+ which adds to Active Model B an additional term that breaks the gradient structure of the active current. This term should be present in general, even if it was not suggested by the microscopic model we chose, which described `quorum sensing' particles. (In fact it can be derived by coarse-graining a model that has both quorum sensing and hard-core collisions; see~\cite{Tjhung18}.)
A careful analysis of the coarsening dynamics of Active Model B+ led us to a generic mechanism whereby bulk phase separation in MIPS is replaced by microphase separation on a finite length-scale. This offers a potentially generic explanation of microphase separation in active matter systems. It also exposes a very interesting theoretical structure in the phase equilibria of stochastic field theories without time-reversal symmetry.

Throughout these notes, we have discussed the emergent behaviour of active field theories primarily, though not exclusively, at the level of mean-field theory. One of the advantages of casting condensed matter problems as field theories is to allow their critical phenomena to be addressed. To do so requires deployment of the renormalization group and other specialised tools. A question arising in that context is whether a given active system has its own critical behaviour or whether it lies in the same universality class as a passive model. This will be discussed elsewhere in the specific context of AMB(+)~\cite{Fernando18}. For related work on non-scalar models, see~\cite{Incomp}. A subtly different question is whether, even if in the same universality class as a passive model, the entropy production of an active field theory scales towards zero (in suitable units) as a given critical point is approached. This is also discussed elsewhere~\cite{Fernando218}.

\section*{Acknowledgements} I thank Irene Li, and also an anonymous student at the School, for critically reading the manuscript; all mistakes are, however, my own.
I acknowledge the contributions of numerous colleagues, students and collaborators with whom we have discussed the topics covered in these notes. An incomplete list is as follows: Fernando Caballero, Suzanne Fielding, Etienne Fodor, Davide Marenduzzo, Cesare Nardini, Sriram Ramaswamy, Alex Solon, Joakim Stenhamar, Julien Tailleur, Adriano Tiribocchi, Elsen Tjhung, Raphael Wittkowski.  I acknowledge funding from the Royal Society in the form of a Research Professorship. I thank Adriano Tiribocchi and Elsen Tjhung for the images in Figures \ref{fig:AMH} and \ref{AMB+phasediag}. Work funded in part by the European Research Council
under the EU's Horizon 2020 Programme, grant number 760769.


\end{document}